\documentclass{emulateapj}

\def\highlight#1{{#1}}

\usepackage{amsmath}
\usepackage{natbib}
\usepackage{soul}

\usepackage{color}
\usepackage{url}

\shorttitle{BROADBAND SPECTRUM AND RADIAL PROFILE OF PWNe}
\shortauthors{ISHIZAKI \textit{ET AL}.}

\begin{document}


\title{Broadband Photon Spectrum and its Radial Profile of Pulsar Wind Nebulae}


\author{Wataru Ishizaki\altaffilmark{1}, Shuta J. Tanaka\altaffilmark{2}, Katsuaki Asano\altaffilmark{1} and Toshio Terasawa\altaffilmark{3}}
%
%


\altaffiltext{1}{Institute for Cosmic Ray Research, The University of Tokyo, 5-1-5 Kashiwa-no-ha, Kashiwa City, Chiba, 277-8582, Japan}
\altaffiltext{2}{Department of Physics, Faculty of Science and Engineering, Konan University, 8-9-1 Okamoto, Kobe, Hyogo, 658-8501, Japan}
\altaffiltext{3}{iTHES Research Group, RIKEN, Wako, Saitama 351-0198, Japan}


\begin{abstract}
The observed radial profiles of the X-ray emission from Pulsar Wind Nebulae (PWNe) have been claimed to conflict with the standard one-dimensional (1-D) steady model.
However, the 1-D model has not been tested to reproduce both the volume-integrated spectrum and the radial profile of the surface brightness, simultaneously.
We revisit the 1-D steady model and apply it to PWNe 3C 58 and G21.5-0.9.
We find that the parameters of the pulsar wind, the radius of the termination shock $r_{\rm s}$ and magnetization $\sigma$, greatly affect both the photon spectrum and radial profile of the emission.
We have shown that the parameters constrained by the entire spectrum lead to a smaller X-ray nebula than observed one.
We have also tested the case that reproduces only the observations in X and gamma-rays, ignoring the radio and optical components.
In this case, there are parameter sets that reproduce both the spectrum and emission profile, but the advection time to the edge of the nebula becomes much smaller than the age.
\highlight{
Our detailed discussion clarifies that the standard 1-D steady model has severe difficulty to reproduce both the volume-integrated spectrum and the surface brightness simultaneously.
This implies that the model should be improved by taking into account extra physical processes such as spatial diffusion of particles.
}
Additionally, we calculate the surface brightness profile of the radio, optical and TeV gamma-rays.
The future observations in these wavelengths are also important to probe the spatial distributions of the relativistic plasma and the magnetic field of PWNe.
\end{abstract}


\section{INTRODUCTION}\label{sec:intro}

Pulsar Wind Nebulae (PWNe) are extended sources around a rotation powered pulsar.
They show a broadband spectrum from radio to $\gamma$-rays so that they would contain very high energy non-thermal particles \citep[]{2006ARAA..44...17G,2009ASSL..357..421K,2014RPPh...77f6901B}.
The central pulsar releases its rotational energy as the pulsar wind,
which is a highly relativistic magnetized electron-positron outflow, and plays a role of the energy source of the PWN.
The strong termination shock, which is formed by the interaction between the pulsar wind and external supernova remnant (or interstellar medium),
has been supposed to produce the non-thermal electrons and positrons, and they emit the synchrotron radiation \citep{1974MNRAS.167....1R}
and the inverse Compton emission \citep{1992ApJ...396..161D}.

PWNe are characterized by a center-filled morphology 
caused by confinement in a supernova remnant (SNR), which is associated with the progenitor of the central pulsar.
Most PWNe are detected as spatially extended sources in radio and X-rays.
While the radio spectral index is almost spatially homogeneous \citep[e.g.,][]{1997ApJ...490..291B,2008MNRAS.386.1411B}, the X-ray spectral index increases with the distance from the pulsar \citep[e.g.,][]{2001AA...376..248B,2004ApJ...616..403S,sch10}.
Recently, PWNe are found to be very bright sources in also TeV $\gamma$-rays \citep{2013arXiv1305.2552K}.
One-zone models of PWNe, which are invented by \citet{1973ApJ...186..249P} and developed by several authors
\citep[e.g.,][]{bed03,che05,2010ApJ...715.1248T,buc11,vor13},
can reproduce broadband spectra of entire nebulae well.
The one-zone models do not describe the spatial distribution of the emission \citep[e.g.,][]{2000AA...359.1107A},
thus it is indispensable to invest a model which includes the spatial structure of nebulae.

The model of \citet[][hereafter KC84s]{1984ApJ...283..694K,1984ApJ...283..710K} has been established as a standard model of PWNe.
They constructed a steady-state and 1-D magnetohydrodynamic model of the Crab Nebula.
Also, assuming the particle acceleration at the termination shock,
they calculated the evolution of non-thermal pairs along the flow and the synchrotron emission from advected particles.
Adopting the model of KC84s, \citet{1996MNRAS.278..525A} succeeded in reproducing the entire photon spectrum including the inverse Compton component.
However, as \citet{Rey03} suggested, it is unclear whether 
KC84s can apply to general PWNe other than the Crab Nebula.
Furthermore, \citet{2004ApJ...616..403S} showed that the model of KC84s disagrees with
the observed radial profile of the X-ray spectral index in 3C 58.
They suggested that the radial profile of the X-ray spectral index in the model should change more rapidly,
and the X-ray nebula size becomes more compact than the observation.
Note that the entire photon spectrum was not taken into account as a model constraint in \citet{2004ApJ...616..403S}.

This ignited the revision of KC84s.
\citet{tan12} introduced the effect of the spatial diffusion of the particles,
and reproduced the X-ray radial profile.
\citet{por16} supported this idea via 3-D magnetohydrodynamic and test-particle simulations.
In each of the studies, simultaneous verification of the entire spectrum and the spatial profile of the emission is not discussed,
thus there is no consensus on the spatial structure in PWN models so far.
In order to advance the study,
it is essential to clarify controversial points in the simple steady 1-D model before introducing nontrivial effects such as the particle diffusion.

We choose two objects, 3C 58 and G21.5-0.9, to examine the 1-D steady model.
Both of the PWNe show the feature that the extent of the X-ray emission is the same as the radio one,
in contrast to the Crab Nebula,
in which the observed size shrinks with increasing frequency.
Furthermore, the 1-D steady model has been never applied to those two PWNe.
Our purpose is to make the validity of the 1-D steady model clear for general PWNe,
so that those two PWNe are suitable targets for testing.

In this paper, we revisit the 1-D steady model and calculate the photon spectrum and its radial profile numerically.
In Section \ref{sec:Model}, we review the 1-D steady model of PWNe based on KC84s.
The parameter dependence of this model is investigated in Section \ref{sec:dependence}.
The application to the two observed sources (3C 58 and G21.5-0.9) is presented in Section \ref{sec:application}.
\highlight{
We discuss the 1-D steady modeling in Section \ref{sec:Discussion} and summarize our results in Section \ref{sec:Conclusion}.
}
\section{Model}\label{sec:Model}
In this paper, 
we adopt the one-dimensional magnetohydrodynamical (MHD) model constructed by KC84s.
We numerically solve the energy distribution of the electron--positron pair plasma along the outflow obtained by MHD equations.
From the resultant spectral distribution of the pairs, we calculate the photon emission from those pairs,
resultant surface brightness, and volume-integrated spectrum.
In this section, we review the the 1-D steady model and present the method to calculate the radial evolution
of the energy distribution of electrons and positrons.

Here, we consider that the PWN is a steady outflow and has a radius $r_{\rm N}$.
The relativistic magnetized wind emitted from the central pulsar forms
a strong termination shock at a radius $r_{\rm s}$.
Assuming that the pre-shock plasma is cold,
the wind property is regulated by three quantities, number density in the comoving frame $n$, bulk Lorentz factor $\gamma$,
and magnetic field in the lab frame $B$ at the shock. 
Almost all of the\textbf{} pulsar spin-down luminosity $L_{\rm sd}$ is converted to the wind luminosity as
\begin{equation} 
L_{\rm sd}=4\pi r_{\rm s}^2 n_{\rm u}u_{\rm u}\gamma_{\rm u}m_{\rm e} c^3 \left(1+\sigma\right),
\label{eq:spindown}
\end{equation}
where $u \equiv \sqrt{\gamma^2-1}$,
the subscript u denotes values at just upstream of the shock,
and $\sigma$ is the ratio of the magnetic energy flux to the particle energy flux at the upstream of the shock,
\begin{equation} 
\sigma\equiv\frac{B_{\rm u}^2/4\pi}{n_{\rm u}u_{\rm u}\gamma_{\rm u}m_{\rm e} c^2}.
\end{equation}
The magnetic field of the wind is dominated by the toroidal component \citep[e.g.,][]{1969ApJ...157..869G}, and the upstream plasma is highly relativistic ($u_{\rm u}/\gamma_{\rm u}\simeq 1$),
which means the downstream temperature is relativistic (adiabatic index $4/3$).
The Rankine-Hugoniot jump conditions provide the values in the downstream (KC84s) as
\begin{equation} 
n_{\rm d}=\frac{n_{\rm u}u_{\rm u}}{u_{\rm d}},
\label{eq:KC_RHa}
\end{equation}
\begin{equation} 
u_{\rm d}^2=\frac{8\sigma^2+10\sigma+1+\sqrt{64\sigma^2\left(\sigma+1\right)^2+20\sigma\left(\sigma+1\right)+1}}{16\left(\sigma+1\right)},
\label{eq:KC_RHb}
\end{equation}
\begin{equation} 
P_{\rm d}=\frac{n_{\rm u}mc^2u_{\rm u}^2}{4\gamma_{\rm d}u_{\rm d}}\left[1+\sigma\left(1-\frac{\gamma_{\rm d}}{u_{\rm d}}\right)\right],
\label{eq:KC_RHc}
\end{equation}
\begin{equation} 
B_{\rm d}=B_{\rm u}\frac{\gamma_{\rm d}}{u_{\rm d}},
\label{eq:KC_RH}
\end{equation}
where the subscript d denotes the values at just downstream of the shock, and $P$ is the thermal pressure.
For $\gamma_{\rm u} \gg 1$ and $\sigma\ll 1$,
we obtain $u_{\rm d}/\gamma_{\rm d} \simeq 1/3$,
which coincides with the well-known result in the relativistic hydrodynamics.

As boundary conditions, we adopt Equations (\ref{eq:KC_RHa})--(\ref{eq:KC_RH}) at the radius $r=r_{\rm s}$,
and solve the steady state and spherical symmetric MHD equations.
Under the toroidal field approximation and the adiabatic assumption,
the MHD equations are integrable.
After some algebra with introducing $\delta \equiv u_{\rm d}/(\sigma u_{\rm u})$, we obtain (KC84s)
\begin{eqnarray} 
\sqrt{1+u^2(r)}
\left(\delta+
\frac{\left(u_{\rm d}^2/\sigma\right)-\frac{1}{2}}{u_{\rm d}^2+\frac{1}{4}}\left(\frac{u(r)r^2}{u_{\rm d}r_{\rm s}^2}\right)^{-\frac{1}{3}}+\frac{u_{\rm d}}{u(r)}\right)
\nonumber \\
=
\gamma_{\rm d}\left(\delta+\frac{\left(u_{\rm d}^2/\sigma\right)-\frac{1}{2}}{u_{\rm d}^2+\frac{1}{4}}+1\right),~~~
\label{eq:KC_vel_eq}
\end{eqnarray}
from which we obtain the radial profile of the four velocity $u(r)$ (or equivalently $\gamma(r)$).
In the strong shock approximation, $u_{\rm d}$ is a function of only $\sigma$ as shown in Equation (\ref{eq:KC_RHb}).
If $\delta \ll 1$ is established,
the above flow Equation (\ref{eq:KC_vel_eq}) is depicted by only one parameter $\sigma$
independently of $n_{\rm u}$ and $u_{\rm u}$.
We calculate the radial profile of $u(r)$ numerically,
while KC84s neglected $\delta$
and adopted $\gamma_{\rm d} \simeq 1$ in the downstream.
Then, the MHD conservation laws provide the other quantities as follows:
\begin{eqnarray}
n_{\rm tot}(r)&=&n_{\rm d}\frac{u_{\rm d}r_{\rm s}^2}{u(r)r^2},\\
\label{eq:beq:ds1}
B(r)&=&B_{\rm d}\frac{\gamma(r)}{\gamma_{\rm d}}\frac{u_{\rm d}r_{\rm s}}{u(r)r},\\
\label{eq:beq:ds2}
P(r)&=&P_{\rm d}\left(\frac{u_{\rm d}r_{\rm s}^2}{u(r)r^2} \right)^{4/3},
\label{eq:beq:ds3}
\end{eqnarray}
where $n_{\rm tot}(r)$ is the comoving number density in the wind.

\begin{figure*}[hbtp]
	\begin{center}
		$\begin{array}{cc}
		\includegraphics[width=0.5\textwidth]{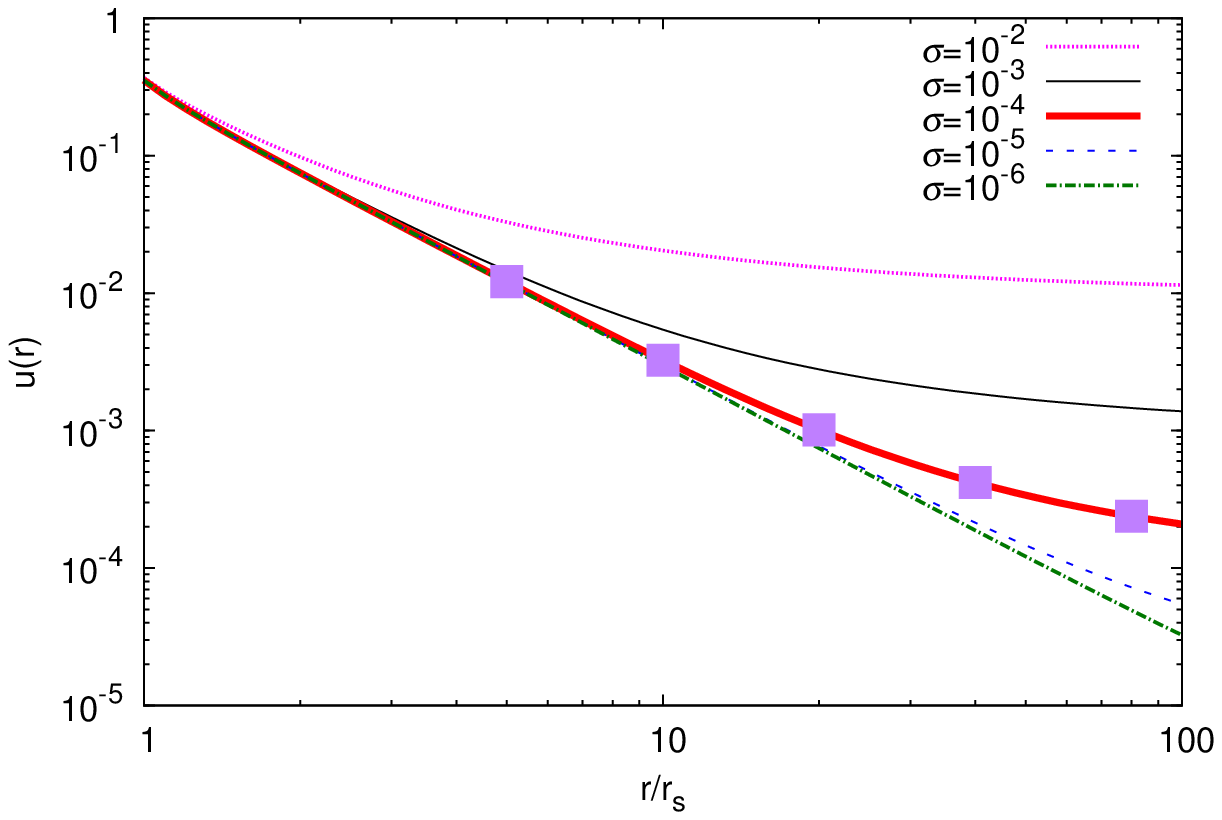} &
		\includegraphics[width=0.5\textwidth]{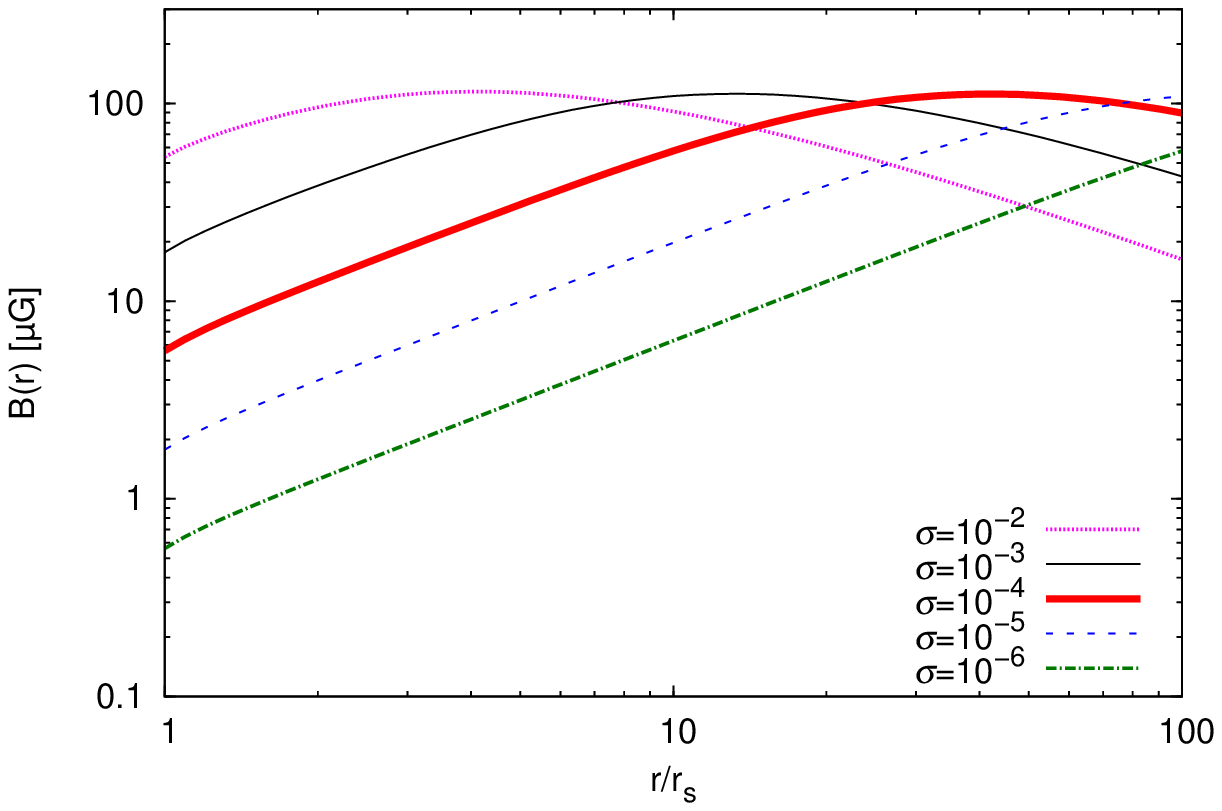}
		\end{array}$
	\end{center}
	\caption{Test calculations of $u$ (left) and $B$ (right) with $u_{\rm u}=10^6$, $L_{\rm sd}=10^{38} \mbox{erg}~\mbox{s}^{-1}$,
and $r_{\rm s}=0.1$ pc for various $\sigma$.
The squares mark values of the four velocity at the edge of the nebula for $r_{\rm N}/r_{\rm s}=5$, $10$, $20$, $40$ and $80$.}
	\label{fig:flow}
\end{figure*}

In Figure \ref{fig:flow}, the radial profiles of $u(r)$ and $B(r)$ in our test calculations
are shown for $\sigma=10^{-6}$, $10^{-5}$, $10^{-4}$, $10^{-3}$ and $10^{-2}$.
The results are almost the same as the behavior shown in KC84s because $\delta \ll 1$ for all the cases.
For $\sigma \ll 1$, at a small radius, the pressure ratio
$\beta_{\rm pl}\equiv B^2/8\pi P$ ($\ll 1$ at $r=r_{\rm s}$) gradually increases with radius as $\beta_{\rm pl} \propto r^2$.
In that regime, the outflow behaves as $u \propto r^{-2}$ (equivalently $n_{\rm tot} \propto r^0$), $B \propto r$ and $P \propto r^0$.
At the radius
\begin{equation}
r\simeq r_{\rm eq} \equiv r_{\rm s}/\sqrt{3\sigma},
\label{eq:req}
\end{equation}
$\beta_{\rm pl}$ becomes unity, namely the magnetic pressure starts to dominate.
Outside this radius, the radial four speed is approximately constant with $B \propto r^{-1}$ and $P \propto r^{-8/3}$.
Thus, the magnetic field has a maximum value at $r\simeq r_{\rm eq}$ as shown in Figure \ref{fig:flow}.

The energy distribution of electron--positron pairs $n(E,r)$ is calculated consistently with the MHD model, 
\begin{equation}
n_{\rm tot}(r)=\int n(E,r)dE.
\label{eq:bPL:con_number}
\end{equation}
At the inner boundary $r=r_{\rm s}$,
$n(E,r_{\rm s})$ is assumed to have
a broken power-law energy distribution at injection as following:
\begin{eqnarray} 
n(E,r_{\rm s})=\left\{
\begin{array}{ll}
\displaystyle\frac{n_0}{E_{\rm b}}\left(\frac{E}{E_{\rm b}}\right)^{-p_1} & \left( E_{\rm min}<E<E_{\rm b} \right) \\
\displaystyle\frac{n_0}{E_{\rm b}}\left(\frac{E}{E_{\rm b}}\right)^{-p_2} & \left( E_{\rm b}<E<E_{\rm max} \right) 
\end{array}
\right.,
\label{eq:bPL:initial}
\end{eqnarray}
where the parameters are the break energy $E_{\rm b}$, minimum energy $E_{\rm min}$, maximum energy $E_{\rm max}$,
and two power-law indices $p_1$ and $p_2$ for low and high energy regions, respectively.
The normalization $n_0=C_1n_{\rm d}$ is determined by Equation (\ref{eq:bPL:con_number}) as
\begin{eqnarray}
C_1 \equiv \left[\frac{1}{p_1-1}\left\{\left(\frac{E_{\rm min}}{E_{\rm b}}\right)^{1-p_1}-1\right\} \right. \nonumber \\
\left. +\frac{1}{p_2-1}\left\{1-\left(\frac{E_{\rm max}}{E_{\rm b}}\right)^{1-p_2}\right\}\right]^{-1}.
\end{eqnarray}
While the origin of the radio spectral component may be different from that for the X-ray and optical components as discussed in KC84s,
in this paper, we use a broken power-law distribution
that has been adopted by the one-zone studies \citep[e.g.,][]{2010ApJ...715.1248T}.

The observed radio spectral index almost uniquely gives the index $p_1$, which is generally lower than 2 \citep[e.g.,][]{1989ApJ...338..171S}.
In this case, the particles with energies $\sim E_{\rm min}$ dominate the particle number.
For simplicity, we fix the minimum energy as $E_{\rm min}=10m_{\rm e}c^2$
and leave the pair density problem \citep[c.f.][]{2013PTEP...12.3E01T}.
The particles above $E_{\rm b}$ may be produced via the shock acceleration \citep{spi08}.
The maximum energy $E_{\rm max}$ is determined by the same method in KC84s; the energy at which a gyro radius is equal to the shock radius provides
\begin{equation}
	E_{\rm max}=eB_{\rm u}r_{\rm s}=\sqrt{ \frac{e^2}{c}\frac{L_{\rm sd}\sigma}{1+\sigma}}.
	\label{eq:ene_max}
\end{equation}
The pressure obtained by $n(E,r_{\rm s})$,
\begin{equation}
P_{\rm d}=\frac{1}{3}\int En(E,r_{\rm s})dE,
\label{eq:bPL:con_pressure}
\end{equation}
should satisfy Equation (\ref{eq:KC_RHc}). We thus obtain
\begin{equation}
n_{\rm d}=\frac{3L_{\rm sd}}{16\pi r_{\rm s}^2cu_{\rm d}\gamma_{\rm d}E_{\rm b}\left(1+\sigma \right)}\frac{C_2}{C_1}\left[1+\sigma\left(1-\frac{\gamma_{\rm d}}{u_{\rm d}}\right)\right],
\label{eq:nd_BPL}
\end{equation}
where
\begin{eqnarray}
C_2=\left[\frac{1}{2-p_1}\left\{1-\left(\frac{E_{\rm min}}{E_{\rm b}}\right)^{2-p_1}\right\} \right. \nonumber \\
\left. +\frac{1}{p_2-2}\left\{1-\left(\frac{E_{\rm max}}{E_{\rm b}}\right)^{2-p_2}\right\}\right]^{-1}.
\end{eqnarray}
Notify that $\gamma_{\rm d}$ or $u_{\rm d}$ is already given
as a function of only $\sigma$ (see Equation (\ref{eq:KC_RHb})), so that the quantities in the upstream are written with the six parameters
$L_{\rm sd}$, $\sigma$, $r_{\rm s}$, $E_{\rm b}$, $p_1$, and $p_2$ as
\begin{equation}
\gamma_{\rm u}=\frac{4}{3}\frac{E_{\rm b}}{mc^2}\gamma_{\rm d}\frac{C_1}{C_2}\left[1+\sigma\left(1-\frac{\gamma_{\rm d}}{u_{\rm d}}\right)\right]^{-1},
\end{equation}
\begin{equation}
B_{\rm u}=\left[\frac{L_{\rm sd}}{cr_{\rm s}^2}\frac{\sigma}{1+\sigma}\right]^{1/2},
\label{eq:mag_us}
\end{equation}
and
\begin{eqnarray}
n_{\rm u}=\frac{9L_{\rm sd}}{64\pi r_{\rm s}^2c \gamma_{\rm d}^2 m_{\rm e} c^2 \left(1+\sigma \right)}\left(\frac{C_2}{C_1}
\frac{m_{\rm e} c^2}{E_{\rm b}} \right)^2 \nonumber \\
\times \left[1+\sigma\left(1-\frac{\gamma_{\rm d}}{u_{\rm d}}\right)\right]^2.
\end{eqnarray}
From the above relations,
the functional shape described in Equation (\ref{eq:bPL:initial})
is also written with the six parameters.

In this 1-D model, we have a unique parameter $r_{\rm s}$,
which is not in the one-zone models,
and $r_{\rm s}$ significantly affects the results as will be shown in Section \ref{sec:dependence}.
The flow solution given by Equation (\ref{eq:KC_vel_eq}) provides the advection time as
\begin{equation}
t_{\rm adv}=\int_{r_{\rm s}}^{r_{\rm N}}\frac{dr}{cu(r)}.
\label{eq:def_tadv}
\end{equation}
While the age of the PWN is an important parameter in the one-zone time-dependent models such as \citet{2010ApJ...715.1248T},
the parameter $r_{\rm s}$ in our steady model adjusts the advection time, 
which may be close to the age of the PWN.

The radial evolution of $n(E,r)$ and photon emission are calculated with the numerical code used in \citet{sas15}, which are based on the time-dependent code
in \citet{asa11} \citep[see also][]{asa12}.
Taking into account the Klein--Nishina effect on the inverse Compton (IC) cooling, the code can follow the temporal evolution
of the energy distribution along the stream.
Transforming the elapsed time into radius as $dr=cu(r) dt$, our calculation is practically equivalent to solve the steady
transport equation (e.g. \citet{1965PSS...13....9P}; \citet{1964ocr..book.....G})
\begin{eqnarray} 
u(r)\frac{\partial n(E,r)}{\partial r}&=&\frac{\partial }{\partial E}\left[ \left(\dot{E}_{\rm syn} +\dot{E}_{\rm IC} \right) n(E,r) \right]
\nonumber \\
&&+\frac{\partial }{\partial E}\left[ \frac{c E n(E,r)}{3r^2}\frac{d}{dr}\left(r^2u(r) \right) \right] \nonumber \\
&&-\frac{c}{r^2}n(E,r)\frac{d}{dr}\left(r^2u(r)\right),
\label{eq:FP}
\end{eqnarray}
where $\dot{E}_{\rm syn}$ and $\dot{E}_{\rm IC}$ are the energy loss rates due to synchrotron radiation and IC scattering, respectively.
The three terms of the right hand side of Equation (\ref{eq:FP}) represent the effects of the radiative cooling, adiabatic cooling and volume expansion, respectively.

Here, we have used the solution for $u(r)$ with adiabatic approximation,
namely the radiative cooling is assumed not to affect the dynamics of the flow.
This approximation is valid when the cooling time for the particles with $E \sim E_{\rm b}$ is longer than the advection time (KC84s).
Most of the results shown in this paper safely satisfy this condition.

\begin{figure}[!htb]
	\centering
	\includegraphics[width=1\linewidth]{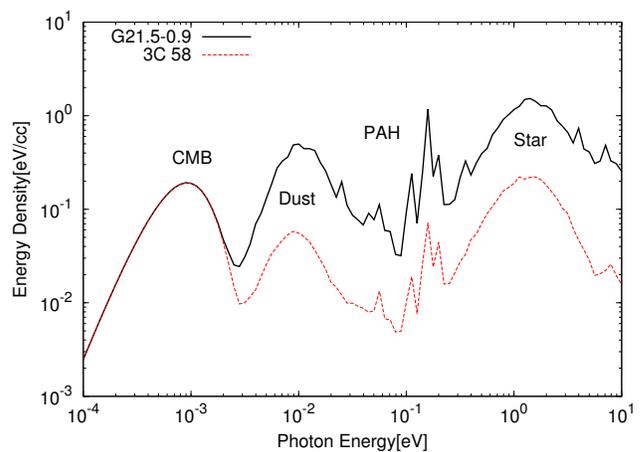}
	\caption[test]{The spectra of the interstellar radiation field taken from GALPROP v54.1.
The black solid line is one adopted for G21.5-0.9, which is located at $R=4\ {\rm kpc}$ and $z=0\ {\rm kpc}$.
The red dashed line is for 3C 58, which is located at $R=9.4\ {\rm kpc}$ and $z=0.5\ {\rm kpc}$.}
	\label{fig:soft_photon}
\end{figure}

The spectral emissivity $j_\nu(r)$ per unit volume is obtained consistently with the energy distribution $n(E,r)$,
the magnetic field profile $B(r)$, and the interstellar radiation field (ISRF) with the Klein--Nishina effect.
The model of the ISRF is taken from GALPROP v54.1 \citep[][and references therein]{vla11}, in which the results of \citet{por05} are adopted,
as shown in Figure \ref{fig:soft_photon}.
The ISRF is assumed as uniform and isotropic in the PWN, and we neglect the synchrotron self-Compton,
which significantly contribute for only limited cases like the Crab Nebula \citep{tor13}.
We neglect the contribution of bremsstrahlung as well, because the density of pairs is low enough \citep{1996MNRAS.278..525A}.

Neglecting the emission from the pre-shock wind, the photon spectrum of the entire nebula $F_{\nu}$ is given by
\begin{equation}
F_{\nu}=\frac{1}{D^2} \int_{r_{\rm s}}^{r_{\rm N}}j_{\nu}\left(r\right) r^2 dr,
\end{equation}
where $D$ is the distance to the PWN from us.
The surface brightness $B_{\nu}$ is also given by
\begin{equation}
B_{\nu}\left (s\right )=2\int_{{\rm max}\left(r_{\rm s},s\right)}^{r_{\rm N}}\frac{j_{\nu}\left(r\right)rdr}{\sqrt{r^2-s^2}},
\end{equation}
where $s$ is the distance perpendicular to the line of sight from the central pulsar.

	Our numerical code is checked by reproducing the result of \citet{1996MNRAS.278..525A} for the Crab Nebula. If we adopt the same assumption as \citet{1996MNRAS.278..525A}, the resultant spectrum agree with the observed one. Note that our more conservative model leads to a slightly dimmer flux for the inverse Compton component as shown in Appendix.
	
\section{Parameter Dependence in the 1-D Steady Model}\label{sec:dependence}
In this section, we discuss how the entire spectrum $L_{\nu}\equiv 4 \pi D^2 F_\nu$
and the surface brightness $B_\nu$ depend on the parameters in this 1-D steady model.
There are some previous studies that discussed the parameter dependence of the 1-D model.
KC84s have already discussed how the parameters $u_{\rm u}$ and $\sigma$ change the total synchrotron luminosity $\int L_\nu d\nu$ (not the spectral distribution).
While \citet{sch10} have studied the X-ray spatial profile for different $r_{\rm s}$
assuming that the flow velocity decreases as a power-law of $r$ independently of $\sigma$,
we investigate the parameter dependence consistently with the MHD flow solution.
We focus on the dependence on the parameters $r_{\rm s}$ and $\sigma$,
which largely affect the spatial structure of the emission.
In this section, the nebula size $r_{\rm N}=2.0\ {\rm pc}$ is fixed.
The external photon field is taken from the model for G21.5-0.9 in Figure \ref{fig:soft_photon}.
For four parameters out of the six parameters in our model, we adopt a parameter set as
$L_{\rm sd}=10^{38}\ {\rm erg}~{\rm s}^{-1}$, $E_{\rm b}=10^5 m_{\rm e} c^2$, $p_1=1.1$, and $p_2=2.5$,
and change $r_{\rm s}$ or $\sigma$ below.

\subsection{Characteristic frequencies and energies of the model}\label{Characteristics}
First, we introduce some typical particle energies and corresponding photon frequencies, and their dependence on the model parameters are discussed.
In this subsection, we limit the discussion within the case of $r_{\rm N}<r_{\rm eq}$ (i.e., $r_{\rm N} / r_{\rm s} <(3\sigma)^{-1/2}$).
In this case, the magnetic field
has a maximum value $3 B_{\rm u} r_{\rm N}/r_{\rm s}$ at the edge of the nebula.
The cooling effect for pairs with energies $E=E_{\rm b}$ is also found to be negligible.
We find the first typical frequency, the intrinsic break frequency,
\begin{eqnarray}
\nu_{\rm b}&=&\frac{3 e B(r_{\rm N})}{4 \pi m_{\rm e} c} \left(\frac{E_{\rm b}}{m_{\rm e} c^2}\right)^2  \nonumber\\
&\simeq& 4.7 \times 10^{12}~\mbox{Hz}
\left( \frac{L_{\rm sd}}{10^{38} \mbox{erg}~\mbox{s}^{-1}} \right)^{\frac{1}{2}}
\left( \frac{\sigma}{10^{-4}} \right)^{\frac{1}{2}} \nonumber \\
&&\times \left( \frac{E_{\rm b}}{10^{5} m_{\rm e} c^2} \right)^{2} 
\left( \frac{r_{\rm s}}{0.1~\mbox{pc}} \right)^{-2}
\left( \frac{r_{\rm N}}{2~\mbox{pc}} \right).
\label{eq:nub}
\end{eqnarray}

In the case of $B \propto r$ or equivalently $u \propto r^{-2}$,
the energy of pairs injected with $E=E_{\rm max}$ decreases with $r$ via the synchrotron cooling as
\begin{equation}
E_{\rm cut}(r)=\frac{E_{\rm max}}{1+\frac{1}{5}\frac{E_{\rm max}}{E_{\rm bof}}\left( (\frac{r}{r_{\rm s}})^5-1\right)},
\label{eq:cut}
\end{equation}
where the burn-off energy
\begin{eqnarray}
E_{\rm bof}&\equiv&\frac{9 m_{\rm e}^4 c^8 u_{\rm d}}{4 e^4 B_{\rm d}^2 r_{\rm s}}\simeq\frac{m_{\rm e}^4 c^9 r_{\rm s}}{8 \sqrt{2}e^4 L_{\rm sd}\sigma} \nonumber\\
&\simeq& 4.3\times10^{17} \mbox{eV}
\left( \frac{L_{\rm sd}}{10^{38} \mbox{erg}~\mbox{s}^{-1}} \right)^{-1} \nonumber\\
&&\times\left( \frac{\sigma}{10^{-4}} \right)^{-1}
\left( \frac{r_{\rm s}}{0.1~\mbox{pc}} \right),
\end{eqnarray}
(KC84s).
At $r=r_{\rm N}$, the maximum energy becomes $E_{\rm cut}(r_{\rm N}) \simeq 5 E_{\rm bof} (r_{\rm s}/r_{\rm N})^5$.
Then, the cooling frequency at the outer boundary is given by
\begin{eqnarray}
\nu_{\rm c}&=&\frac{3 e B(r_{\rm N})}{4 \pi m_{\rm e} c} \left(\frac{E_{\rm cut}(r_{\rm N})}{m_{\rm e} c^2}\right)^2  \nonumber\\
&\simeq& 8.2 \times 10^{14}~\mbox{Hz}
\left( \frac{L_{\rm sd}}{10^{38} \mbox{erg}~\mbox{s}^{-1}} \right)^{-\frac{3}{2}}
\left( \frac{\sigma}{10^{-4}} \right)^{-\frac{3}{2}} \nonumber \\
&&\times \left( \frac{r_{\rm s}}{0.1~\mbox{pc}} \right)^{10}
\left( \frac{r_{\rm N}}{2~\mbox{pc}} \right)^{-9}.
\label{eq:nuc}
\end{eqnarray}
Above $\nu_{\rm c}$, the entire spectrum should show the softening behavior.

The maximum particle energy decreases following Equation (\ref{eq:cut}),
while the magnetic field increases as $B(r)=3 B_{\rm u} (r/r_{\rm s})$.
The typical synchrotron frequency $\propto B(r) E_{\rm cut}(r)^2$ peaks at
\begin{eqnarray}
r \simeq r_{\rm pk} &\equiv& \left( \frac{5 E_{\rm bof}}{9 E_{\rm max}} \right)^{1/5} r_{\rm s} \nonumber\\
&\simeq& 0.14 {\rm pc}
\left( \frac{L_{\rm sd}}{10^{38} \mbox{erg}~\mbox{s}^{-1}} \right)^{-3/10} \nonumber\\
&&\times \left( \frac{\sigma}{10^{-4}} \right)^{-3/10} 
\left( \frac{r_{\rm s}}{0.1~\mbox{pc}} \right)^{6/5},
\end{eqnarray}
where we have assumed $E_{\rm max} \ll 5 E_{\rm bof}$.
With $r_{\rm pk}$, the cut-off frequency in the synchrotron spectrum is obtained as
\begin{eqnarray}
\nu_{\rm cut}&=&\frac{3 e B(r_{\rm pk})}{4 \pi m_{\rm e} c} \left(\frac{E_{\rm cut}(r_{\rm pk})}{m_{\rm e} c^2}\right)^2  \nonumber\\
&\simeq& \frac{729 e}{400 \pi m_{\rm e} c}
\sqrt{\frac{L_{\rm sd} \sigma}{c r_{\rm s}^2}} \left( \frac{5 E_{\rm bof}}{9 E_{\rm max}} \right)^{1/5}
\left(\frac{E_{\rm max}}{m_{\rm e} c^2}\right)^2 \nonumber \\
&\simeq& 9.3 \times 10^{18}~\mbox{Hz}
\left( \frac{L_{\rm sd}}{10^{38} \mbox{erg}~\mbox{s}^{-1}} \right)^{11/10} \nonumber \\
&&\times \left( \frac{\sigma}{10^{-4}} \right)^{11/10} 
\left( \frac{r_{\rm s}}{0.1~\mbox{pc}} \right)^{-4/5},
\label{eq:nucut}
\end{eqnarray}
above which the flux decreases exponentially.

\subsection{Shock radius dependence}
\begin{figure*}[!htb]
	\begin{center}
		$\begin{array}{cc}
		\includegraphics[width=0.5\textwidth]{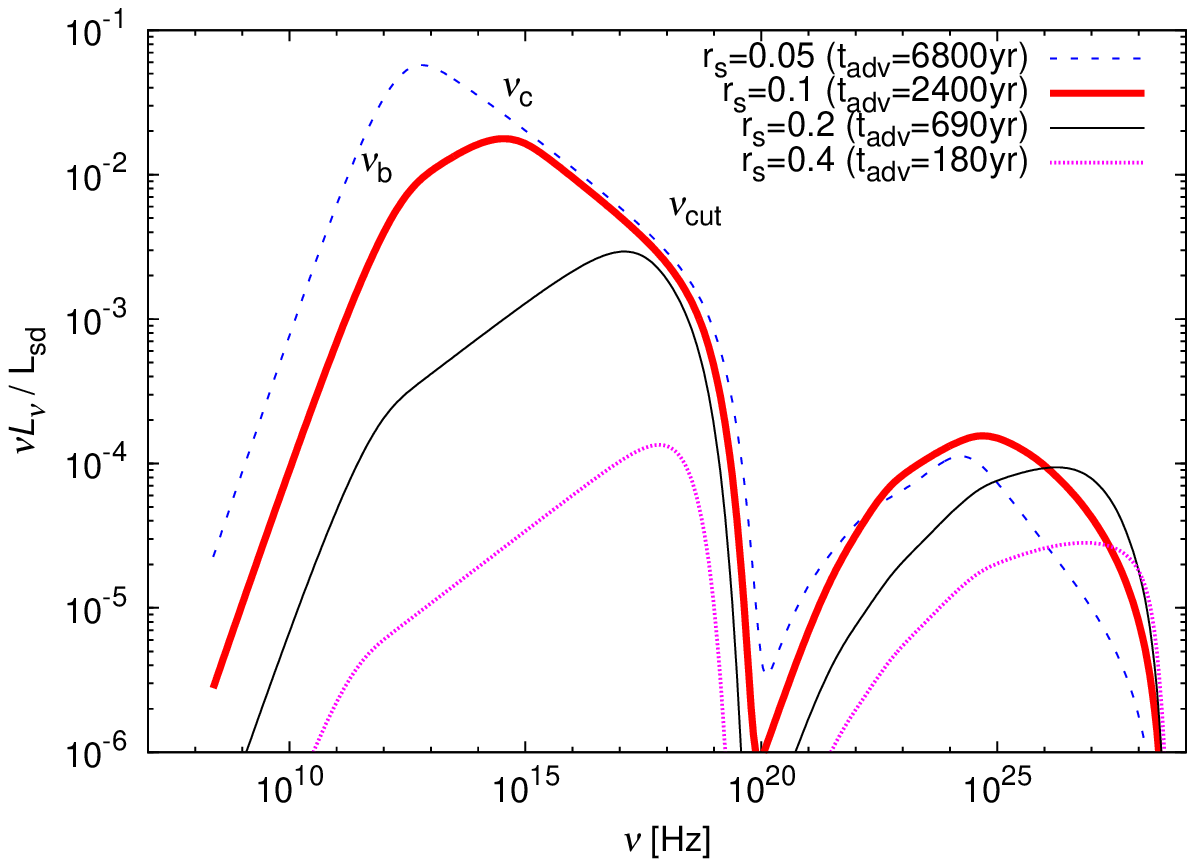} &
		\includegraphics[width=0.5\textwidth]{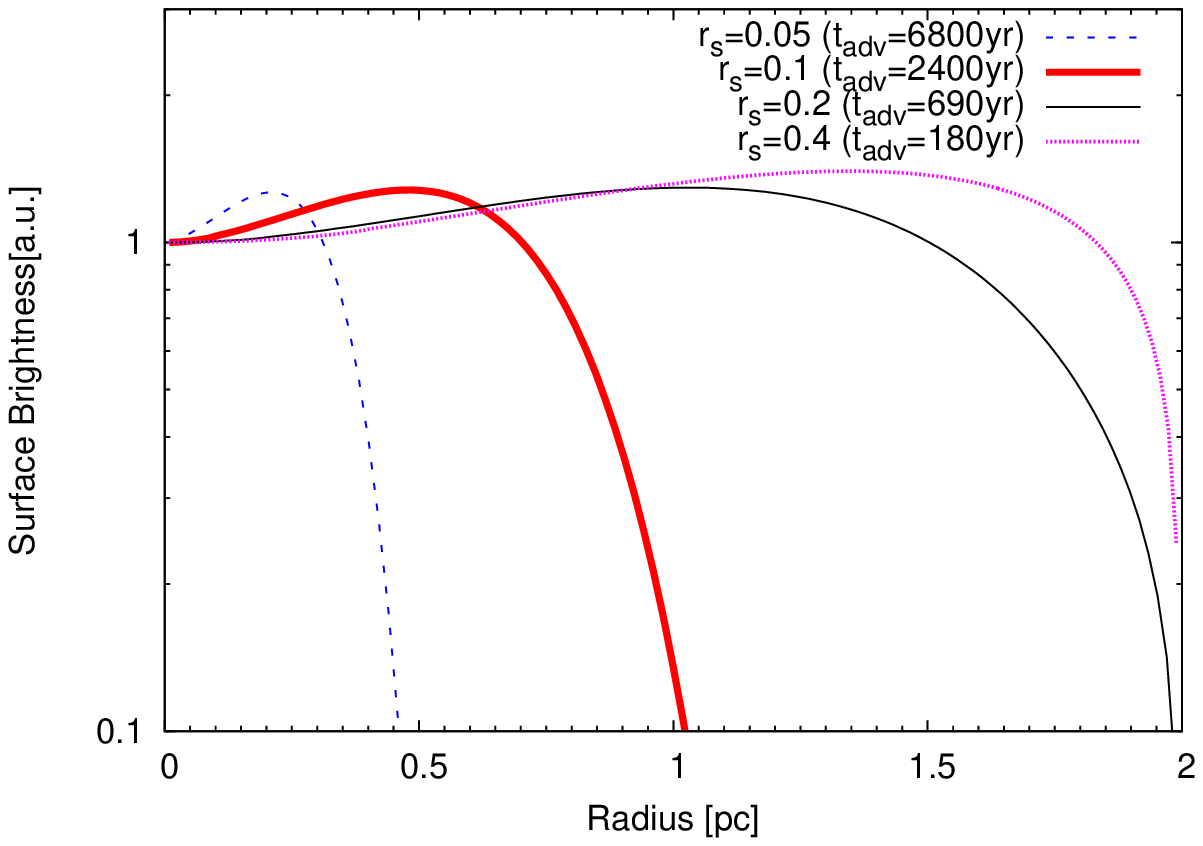}
		\end{array}$
	\end{center}
	\caption{Test calculations to see the shock radius dependence. The parameters are $L_{\rm sd}=10^{38}\ {\rm erg}~{\rm s}^{-1}$, $E_{\rm b}=10^5 m_{\rm e} c^2$, $\sigma=10^{-4}$,
$p_1=1.1$, and $p_2=2.5$.
(Left) The entire spectrum calculated for various values of $r_{\rm s}$ (see Table \ref{rs_dep_list}).
(Right) The radial profile of the X-ray surface brightness for 0.5-10 keV range. The nebula radius $r_{\rm N}$ is $2$ pc.}
	\label{fig:rs_dependence}
\end{figure*}

One-zone models do not include the shock radius as a parameter.
The value $r_{\rm s}$ is a characteristic parameter in the 1-D model.
For the Crab Nebula \citep[][]{sch13}, Vela \citep{hel01}, and MSH 15-52 \citep{yat09},
a possible shock structure (inner ring) is detected with X-ray observations.
In most of PWNe, however, the shock radii are not observationally constrained well.
In Figure \ref{fig:rs_dependence}, we show the $r_{\rm s}$-dependences of the entire spectrum $L_{\nu}$ and the X-ray surface brightness with $\sigma=10^{-4}$.
The synchrotron component dominates below $10^{20}$ Hz ($\sim 400$ keV) and the IC component dominates in $\gamma$-rays.
Note that in the case for $r_{\rm s}=0.05$ pc, 
$E_{\rm cut}(r)<E_{\rm b}$ beyond $ r\gtrsim 1.5$ pc
, i.e., the adiabatic approximation is invalid.
In Table \ref{rs_dep_list}, for various $r_{\rm s }$, we summarize the advection time, the volume-averaged magnetic field $B_{\rm av}$ as,
\begin{equation}
\frac{B_{\rm av}^2}{8\pi}=\int_{r_{\rm s}}^{r_{\rm N}}\frac{B(r)^2}{8\pi}4\pi r^2dr\bigg/\int_{r_{\rm s}}^{r_{\rm N}}4\pi r^2dr,
\end{equation}
and the maximum magnetic field $ B_{\rm max} =B(r_{\rm N})$ because we have $ r_{\rm eq} >r_{\rm N}$ from Equation (\ref{eq:req}) with $ \sigma=10^{-4} $.

First, as a benchmark case, we take up the case of $r_{\rm s}=0.1$ pc (the red solid line in Figure \ref{fig:rs_dependence}), which leads to $ r_{\rm eq}=5.7$ pc.
There are two breaks in the synchrotron spectrum: the intrinsic break at $\nu_{\rm b} \sim 2.2 \times 10^{12}$ Hz
corresponding to $E_{\rm b}$ and the cooling break at $\nu_{\rm c} \sim 3.6 \times 10^{14}\ {\rm Hz}$.
Adopting the average magnetic field and the advection time, the cooling break energy of pairs is
\begin{equation}
E_{\rm c}^{({\rm av})}\simeq \frac{6 \pi m_{\rm e}^2 c^3}{\sigma_{\rm T} B_{\rm av}^2 t_{\rm adv}}\simeq 670~\mbox{GeV}
\left( \frac{B_{\rm av}}{88 \mu\mbox{G}} \right)^{-2} \left( \frac{t_{\rm adv}}{2400 \mbox{yr}} \right)^{-1}.
\label{eq:Ec_ave}
\end{equation}
The corresponding cooling break frequency is written as
\begin{eqnarray}
\nu_{\rm c}^{({\rm av})}&=&\frac{3 e B_{\rm av}}{4 \pi m_{\rm e} c} \left(\frac{E_{\rm c}^{\rm (av)}}{m_{\rm e} c^2}\right)^2  \nonumber\\
&\simeq& 6.4 \times 10^{14}~\mbox{Hz}
\left( \frac{B_{\rm av}}{88 \mu\mbox{G}} \right)^{-3} \left( \frac{t_{\rm adv}}{2400 \mbox{yr}} \right)^{-2}.
\label{eq:nuc_ave}
\end{eqnarray}
The above value obtained with a one-zone-like treatment roughly agree with our results.

\begin{table*}[!hbtp]
	\begin{center}
		\begin{tabular}{lrrrr}
			\hline\hline
			Radius of termination shock (pc) & $0.05$ & $0.1$ & $0.2$ & $0.4$ \\
			\hline
			Total number of pairs ($10^{50}$) & $9.5$ & $3.4$ & $0.97$ & $0.25$ \\
			Advection time (yr) & $6800$ & $2400$ & $690$ & $180$ \\
			Total pressure at $r=r_{\rm N}$ ($10^{-10}$dyn~$\mbox{cm}^{-2}$) & $36$ & $16$ & $5.4$ & $1.5$ \\
			Maximum magnetic field ($\mu$G) & $250$ & $100$ & $32$ & $8.7$ \\
			Averaged magnetic field ($\mu$G) & $240$ & $88$ & $26$ & $6.8$ \\
			$r_{\rm eq}/r_{\rm N}$ & $1.4$ & $2.9$ & $5.8$ & $12$ \\
			\hline\hline
		\end{tabular}
	\end{center}
	\caption{Obtained parameters for the test calculation with $\sigma=10^{-4}$ shown in Fig. \ref{fig:rs_dependence}.
See Fig. \ref{fig:rs_dependence} for the other parameters.
In these parameter sets, the magnetic field always reaches its maximum at $r=r_{\rm N}$}
	\label{rs_dep_list}
\end{table*}

The analytical descriptions of the spectral indices $\alpha$ ($F_\nu \propto \nu^{-\alpha}$) are $(p_1-1)/2$ below $\nu_{\rm b}$,
and $(p_2-1)/2$ for $\nu_{\rm b}<\nu<\nu_{\rm c}$. The calculated spectrum agrees with those values.
In one-zone models, the index above $\nu_{\rm c}$ steepened by $1/2$.
However, in the 1-D models following $B \propto r$ and $u \propto r^{-2}$,
the spectral change $\Delta=(p+7)/18$ is slightly different from 1/2 \citep{2009ApJ...703..662R},
which also agrees with our result.

Let us remark the flux at $\nu=\nu_{\rm cut}$.
When the synchrotron cooling is efficient for particles of $E=E_{\rm max}$,
almost all of those energies are released by photon emission until $r=r_{\rm pk}$.
Since the energy density of pairs, which have an energy $ E=E_{\rm max} $ at $ r=r_{\rm s} $, is estimated as $L_{\rm sd}(E_{\rm max}/E_{\rm b})^{2-p_2}$, 
the synchrotron luminosity at $\nu\sim\nu_{\rm cut}$ is calculated as $\nu_{\rm cut}L_{\nu_{\rm cut}} \sim 1.7\times10^{-2} L_{\rm sd}$ for $\sigma=10^{-4}$ and $p_2=2.5$,
which seems consistent with the flux in Figure \ref{fig:rs_dependence}.
The above estimate does not depend on $r_{\rm s}$, which agrees with the results for $r_{\rm s}\leq 0.2$ pc, where $\nu_{\rm c}<\nu_{\rm cut}$.

A particle of energy $E$ emits synchrotron photons of frequency $\nu \propto E^2 B$, and power $ p_{\rm syn}\propto E^2 B^2 $, 
then the spectral emissivity $j_\nu$ is proportional to $n(E) E^2 B^2 (dE/\nu)$.
When we can assume $\sigma \ll 1$ and $n_{\rm tot} \propto r^0$,
$j_\nu \propto n_0 E_{\rm b}^{p-1} \nu^{-(p-1)/2} B(r)^{(p+1)/2}$ ($p$ is index of the pair energy distribution) at the energy range where the cooling effect is negligible.
The constant density implies that the magnetic field behaves as $B \sim B_{\rm d} r/r_{\rm s}
\propto L_{\rm sd}^{1/2} r_{\rm s}^{-2} \sigma^{1/2} r$.
Since we can treat $C_2$ as a constant for $p_1 < 2 < p_2$ and $E_{\rm min}\ll E_{\rm b} \ll E_{\rm max}$,
we obtain $n_0 \propto L_{\rm sd} r_{\rm s}^{-2} E_{\rm b}^{-1}$.
Finally, we obtain the entire specific luminosity ($L_{\rm \nu} \sim 4 \pi \int_{r_{\rm s}}^{r_{\rm N}} dr r^2 j_\nu$) as
\begin{equation}
\nu L_\nu\propto L_{\rm sd}^{\left(p+5\right)/4} E_{\rm b}^{p-2} \sigma^{\left(p+1\right)/4}r_{\rm s}^{-(p+3)}
r_{\rm N}^{(p+7)/2}\nu^{-(p-3)/2},
\label{eq:nulnu}
\end{equation}
where $p$ is $p_1$ for $\nu < \nu_{\rm b}$, and $p_2$ for $\nu_{\rm b}< \nu < \nu_{\rm c}$.
At $\nu=\nu_{\rm b} \propto L_{\rm sd}^{1/2} \sigma^{1/2} E_{\rm b}^2 r_{\rm s}^{-2} r_{\rm N}$ (Equation (\ref{eq:nub})),
\begin{eqnarray}
\frac{\nu_{\rm b} L_{\nu_{\rm b}}}{L_{\rm sd}}&\simeq& 7 \times 10^{-3}
\left( \frac{L_{\rm sd}}{10^{38} \mbox{erg}~\mbox{s}^{-1}} \right)
\left( \frac{\sigma}{10^{-4}} \right)\nonumber \\
&&\times \left( \frac{E_{\rm b}}{10^5 m_{\rm e} c^2} \right)
\left( \frac{r_{\rm s}}{0.1~\mbox{pc}} \right)^{-6}
\left( \frac{r_{\rm N}}{2~\mbox{pc}} \right)^{5}.
\label{eq:nulnum}
\end{eqnarray}
where the absolute value, which is difficult to evaluate analytically, is provided by our numerical results.
Above $\nu=\nu_{\rm c} \propto L_{\rm sd}^{-3/2} \sigma^{-3/2} r_{\rm s}^{10} r_{\rm N}^{-9}$ (Equation (\ref{eq:nuc})),
using the above formula,
the spectrum behaves as $\nu L_\nu \sim \nu_{\rm c} L_{\nu_{\rm c}} (\nu/\nu_{\rm c})^{-5 (p_2-2)/9}$
so that
\begin{equation}
\nu L_\nu\propto L_{\rm sd}^{\left(p_2+4\right)/6} E_{\rm b}^{p_2-2} \sigma^{\left(p_2-2\right)/6}r_{\rm s}^{-4(p_2-2)/9}\nu^{-5(p_2-2)/9},
\label{eq:analytical:l2}
\end{equation}
for $\nu_{\rm c}<\nu<\nu_{\rm cut}$.
Those formulae well agree with our numerical results.

The cooling frequency strongly depends on $r_{\rm s}$.
The frequencies $\nu_{\rm b}$ and $\nu_{\rm c}$ have similar values for $r_{\rm s}=0.05$ pc,
while $\nu_{\rm c}$ and $\nu_{\rm cut}$ merge for $r_{\rm s}=0.2$ pc.
For $r_{\rm s}=0.4$ pc, $\nu_{\rm c}$ becomes higher than $\nu_{\rm cut}$.
Namely, both of the radiative and adiabatic cooling effects are negligible in this case.
This leads to the lower flux at $\nu=\nu_{\rm cut}$ for $r_{\rm s}=0.4$ pc than the fluxes for smaller $r_{\rm s}$.

The value of $E_{\rm cut}(r_{\rm N})$ increases with $r_{\rm s}$, because the synchrotron cooling becomes less effective.
Reflecting this, the IC spectra show a soft-to-hard evolution with $r_{\rm s}$.
The high-energy cut-off of the IC component is determined by the maximum energy of pairs.
Electron--positron pairs expend only a small fraction of their energy in the IC emission.
Since the $L_{\rm sd}$ and $E_{\rm b}$ are common for the examples in Figure \ref{fig:rs_dependence},
the IC flux in the low energy range ($10^{20}-10^{22} {\rm Hz}$) is basically proportional to the total number of corresponding low-energy particles in the nebula.
As shown in the left panel in Figure \ref{fig:flow}, a flow with a small ratio of $r_{\rm N}/r_{\rm s}$
reaches the edge of the nebula before significant deceleration.
Consequently, the advection time becomes shorter for a smaller $r_{\rm N}/r_{\rm s}$ as shown in Table \ref{rs_dep_list}.
If we can neglect the cooling effect, the total particle number $\propto t_{\rm adv} L_{\rm sd}E_{\rm b}^{p_1-2}$ decreases with $r_{\rm s}$.
This effect is seen as the flux growth with increasing $t_{\rm adv}$ below $\sim 10^{22}$ Hz.
For $r_{\rm s}=0.05$ pc, the synchrotron cooling is crucial ($E_{\rm cut}(0.7r_{\rm N})\lesssim E_{\rm b}$),
which practically reduces the particle number above $E_{\rm b}$ in the nebula.
Therefore, the IC flux in this case does not follow the aforementioned trend of the flux growth.

The surface brightness profile in X-rays (see the right panel of Figure \ref{fig:rs_dependence}) is
regulated by the synchrotron cooling. For a smaller $r_{\rm s}$, the stronger magnetic field
results in a compact X-ray profile.
The X-ray extent is proportional to $r_{\rm s}^{10/9}$ for $r_{\rm s}\leq 0.2$ pc as explained as follows.
When the cooling effect is significant, $E_{\rm cut} \propto E_{\rm bof} (r_{\rm s}/r)^5 \propto \sigma^{-1} r_{\rm s}^6 r^{-5}$,
while the magnetic field behaves as $B \propto \sigma^{1/2} r_{\rm s}^{-2} r$.
For a given frequency $\nu \propto B E^2$, the maximum radius to emit photons of $\nu$
is proportional to $\sigma^{-1/6} r_{\rm s}^{10/9}$. This supports the X-ray size growth with $r_{\rm s}^{10/9}$
for the case where the cooling effect is significant.
For $r_{\rm s}=0.4$ pc, the synchrotron cooling is negligible effect for the X-ray profile.
In order to reconcile the fact that the X-ray extent is comparable to the radio nebula,
a large $r_{\rm s}$ is preferable, though the synchrotron component becomes very hard and dim in this case.

As shown in Table \ref{rs_dep_list}, the total pressure $P_{\rm tot} \equiv 4P u^2+P+B^2/8\pi$ at the outer boundary,
which may balance with the pressure outside the nebula,
decreases with $r_{\rm s}$ by roughly an order of magnitude.
Since the uncertainty of the current observation of the outside pressure is larger than this variance,
it may be difficult to constrain the value of $r_{\rm s}$ directly.

\subsection{$\sigma$ dependence}\label{dep:sig}
\begin{figure*}[!htb]
	\begin{center}
		$\begin{array}{cc}
		\includegraphics[width=0.5\textwidth]{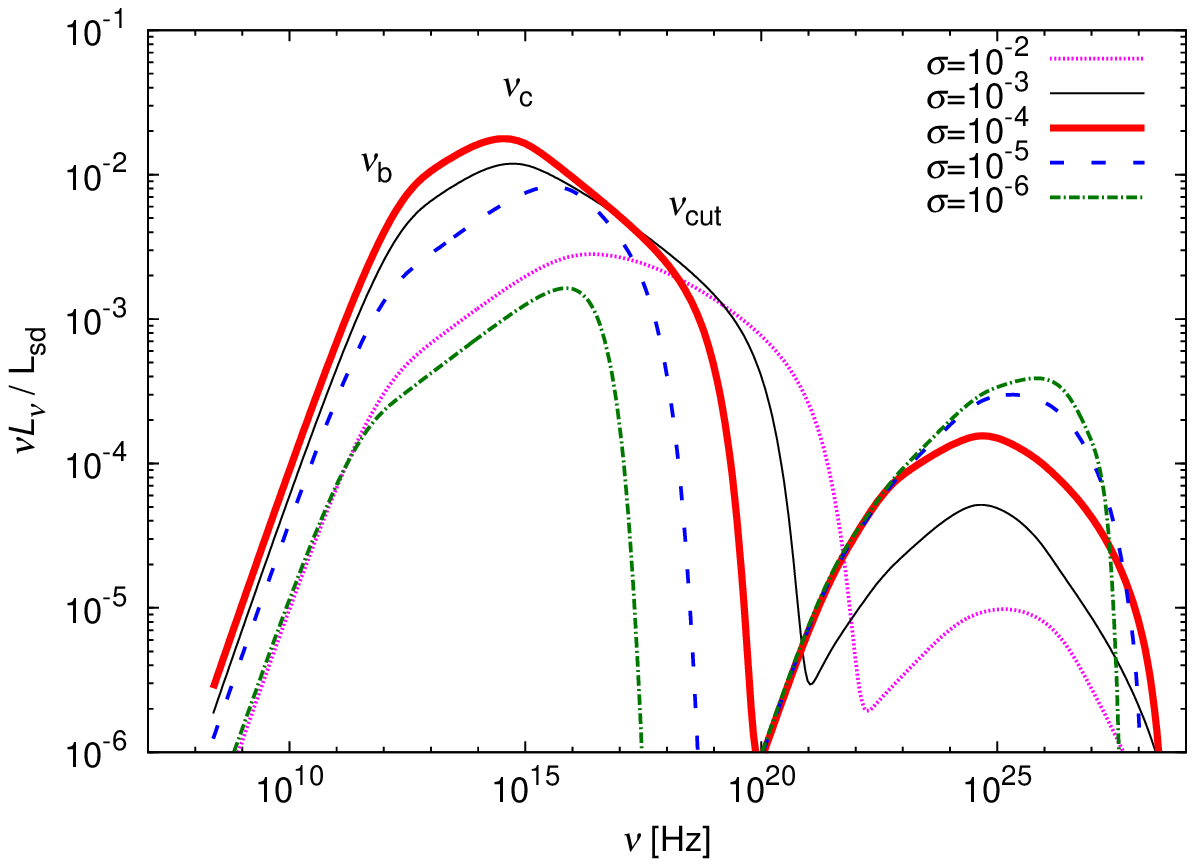} &
		\includegraphics[width=0.5\textwidth]{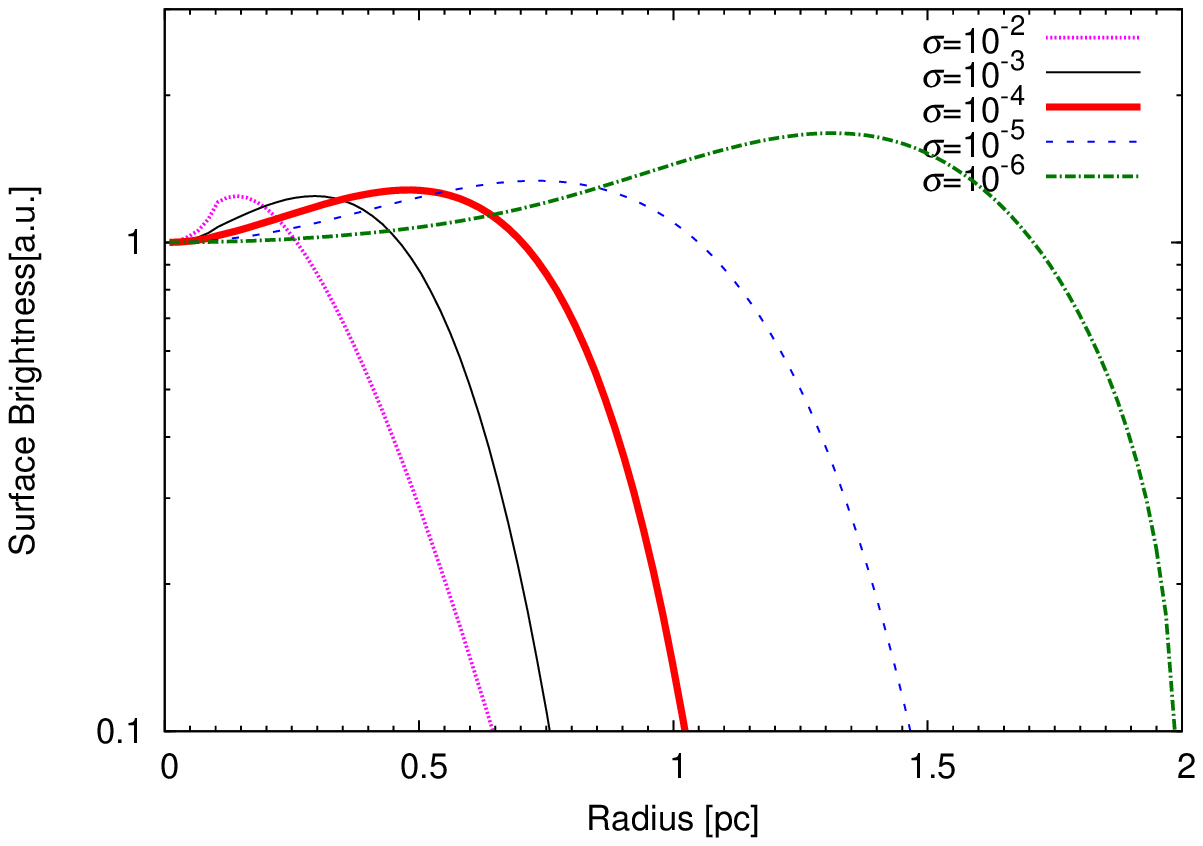}
		\end{array}$
	\end{center}
	\caption{Test calculations to see $\sigma$ dependence. The parameters are the same as those in Fig. \ref{fig:rs_dependence}
except for $\sigma$ and $r_{\rm s}=0.1$ pc.
(Left) The entire spectrum calculated for various values of $\sigma$ (see Table \ref{sigma_dep_list}).
(Right) The radial profile of the X-ray surface brightness for 0.5-10 keV range. }
	\label{fig:sigma_dependence}
\end{figure*}

Figure \ref{fig:sigma_dependence} shows the $\sigma$ dependences of
the entire spectrum and the X-ray surface brightness.
While the shock radius is fixed to $r_{\rm s}=0.1$ pc and the other parameters are the same as those in the previous subsection,
the value of $\sigma$ changes from $10^{-6}$ to $10^{-2}$.
As is shown in the figure, a complicated behavior appears in the spectral shape with increasing $\sigma$.
The change of $\sigma$ modifies the profiles of the emission through two processes: 
the strength of magnetic field (see Equation (\ref{eq:mag_us})) and the deceleration profile as shown in Fig. \ref{fig:flow}.
The magnetic field strength affects the typical synchrotron frequency and the cooling efficiency.
The flow velocity profile adjusts the radius $r_{\rm eq}$, where the magnetic field becomes maximum,
and the advection time, which controls the total energy in the nebula
and the cooling efficiency (the ratio of the cooling time to the advection time).

For $\sigma \leq 10^{-4}$, the radius $r_{\rm eq}$ is outside $r_{\rm N}$
(see Table \ref{sigma_dep_list}, note that $r_{\rm N}/r_{\rm s}$ corresponds to 20 in Figure \ref{fig:flow}),
so that the behaviors of the characteristic frequencies
are well explained by Equations (\ref{eq:nub}),
(\ref{eq:nuc}), and (\ref{eq:nucut}) as $\nu_{\rm b} \propto \sigma^{1/2}$, $\nu_{\rm c} \propto \sigma^{-3/2}$,
and $\nu_{\rm cut} \propto \sigma^{11/10}$.
In the case of $\sigma=10^{-6}$, the frequency $\nu_{\rm c}$ is much higher than $\nu_{\rm cut}$,
so that the power-law portion for $\nu_{\rm c}<\nu<\nu_{\rm cut}$ is absent.
Below $\nu_{\rm cut}$, the formulae of Equations (\ref{eq:nulnu}) and (\ref{eq:nulnum})
well represent the spectral behavior for $\sigma\leq 10^{-4}$.
In the frequency range of $\nu_{\rm c}<\nu<\nu_{\rm cut}$,
the spectrum practically follows
Equation (\ref{eq:analytical:l2}) for the cases
of $\sigma=10^{-5}$ and $10^{-4}$.
As $\sigma$ increases, the resultant stronger magnetic field leads to
a large efficiency of the energy release during the advection time.
The peak of the  synchrotron flux at $\nu=\nu_{\rm c}$ grows with $\sigma^{p_2-2}$ 
(see Equations (\ref{eq:nuc}) and (\ref{eq:nulnu}))
for $\sigma \leq 10^{-4}$, accompanying the shift of $\nu_{\rm c}$ to a lower frequency.

\begin{table*}[!hbtp]
	\begin{center}
		\begin{tabular}{lrrrrr}
			\hline\hline
			Magnetization Parameter & $10^{-2}$ & $10^{-3}$ & $10^{-4}$ & $10^{-5}$ & $10^{-6}$\\
			\hline
			Total number of pairs ($10^{50}$) & $0.37$ & $1.6$ & $3.4$ & $4.0$ & $4.2$\\
			Advection time (yr) & $280$ & $1200$ & $2400$ & $2900$ & $2900$\\
			Total pressure at $r=r_{\rm N}$ ($10^{-10}$erg~cm$^{-3}$) & $2.3$ & $8.9$ & $16$ & $21$ & $23$ \\
			Maximum magnetic field ($\mu$G) & $130$ & $130$ & $100$ & $43$ & $14$\\
			Magnetic field at $r=r_{\rm N}$ ($\mu$G) & $69$ & $120$ & $100$ & $43$ & $14$\\
			Averaged magnetic field ($\mu$G) & $86$ & $120$ & $87$ & $34$ & $11$\\
			$r_{\rm eq}/r_{\rm N}$ & $0.29$ & $0.91$ & $2.89$ & $9.13$ & $28.9$ \\
			\hline\hline
		\end{tabular}
	\end{center}
	\caption{Obtained parameters for the test calculation with $r_{\rm s}=0.1$ pc shown in Fig. \ref{fig:sigma_dependence}.
See Fig. \ref{fig:rs_dependence} for the other parameters.}
	\label{sigma_dep_list}
\end{table*}

The spectral behavior
deviates from the above trends for $\sigma \geq 10^{-3}$.
This is because the radial evolution of the magnetic field
can be no longer approximated by $B \propto r$.
As shown in Table \ref{sigma_dep_list},
the radius $r_{\rm eq}$ is inside the nebula radius $r_{\rm N}$ in this parameter region unlike the discussion in Section \ref{Characteristics}.
The magnetic pressure prevents the flow from deceleration at $r>r_{\rm eq}$,
and the adiabatic cooling starts to play its role.
The decline of the advection time (see Table \ref{sigma_dep_list})
leads to the reduction of the total energy in the nebula.
As a result, above $\sigma=10^{-4}$, the synchrotron peak flux turns into decreasing
and the cooling frequency turns into increasing.
Therefore, we cannot decrease $\nu_{\rm c}$ extremely low in the 1-D steady model differently from the one-zone models.

In the case of $r_{\rm eq}<r_{\rm N}$, the contribution to the entire spectrum is dominated by the emission from non-thermal pairs at $r<r_{\rm eq}$.
At $r>r_{\rm eq}$, electrons/positrons lose their energies via adiabatic cooling
as $E \propto (r/r_{\rm eq})^{-2/3}$ and the magnetic field decays as $\propto r^{-1}$.
Thus, the energy loss rate rapidly decreases as $r^{-10/3}$ so that
the emission beyond $r=r_{\rm eq}$ is almost negligible.
From Equation (\ref{eq:cut}), assuming $E_{\rm max} \gg E_{\rm bof}$,
we obtain $E_{\rm cut}(r_{\rm eq})\simeq 5 E_{\rm bof} (r_{\rm s}/r_{\rm eq})^5 \simeq 5 E_{\rm bof} (3 \sigma)^{5/2}$.
In this case, the cooling frequency for enough large $\sigma$ that $r_{\rm eq}$ is smaller than $r_{\rm N}$ may be estimated with $E_{\rm cut}$ at $r=r_{\rm eq}$ as
\begin{eqnarray}
\nu^{{\rm (eq)}}_{\rm c}&=&\frac{3 e B(r_{\rm eq})}{4 \pi m_{\rm e} c} \left(\frac{E_{\rm cut}(r_{\rm eq})}{m_{\rm e} c^2}\right)^2  \nonumber\\
&\simeq& 5.9 \times 10^{16}~\mbox{Hz}
\left( \frac{L_{\rm sd}}{10^{38} \mbox{erg}~\mbox{s}^{-1}} \right)^{-\frac{3}{2}} \nonumber \\
&&\times\left( \frac{\sigma}{10^{-2}} \right)^{3} 
\left( \frac{r_{\rm s}}{0.1~\mbox{pc}} \right). 
\label{eq:nucsig}
\end{eqnarray}
Although the estimate of the value of $\nu^{\rm (eq)}_{\rm c}$ almost agrees with our numerical results,
the dependence on $\sigma$ is a little bit strong.
This is because the case for $\sigma=10^{-2}, 10^{-3}$ are in a marginal situation ($r_{\rm eq} \sim r_{\rm N}$).
Nevertheless, it is certain that the cooling frequency for $\sigma>10^{-4}$ ($r_{\rm eq}<r_{\rm N}$) becomes large with increasing $\sigma$.

In the frequency range of $\nu^{\rm (eq)}_{\rm c}<\nu<\nu_{\rm cut}$ for a larger $\sigma$,
the spectrum is harder than the analytical estimate $\alpha=(5 p_2-1)/9$ for $r_{\rm eq}<r_{\rm N}$.
On the other hand, since $r_{\rm pk}\sim r_{\rm s}$,
the frequency $\nu_{\rm cut}$ well agrees with the analytical estimate
of Equation (\ref{eq:nucut}) even for a larger $\sigma$.
For $\sigma \geq 10^{-4}$,
as we have mentioned in the previous subsection, the luminosity around $\nu_{\rm cut}$
is almost independent of $\sigma$,
while the peak luminosity at $\nu=\nu_{\rm c}$ decreases with increasing $\sigma$
following the decline of the total energy in the nebula.
Those complicated effects lead to the spectral hardening between $\nu_{\rm c}$ and $\nu_{\rm cut}$.

The peak flux of the IC component declines monotonically with increasing $\sigma$.
Below the spectral break frequency $10^{23}$ Hz, which corresponds
to the photon energy emitted
by particles of $E_{\rm b}$ interacting with dust photons,
all the model curves for $\sigma \leq 10^{-4}$
almost overlap each other.
In this range of $\sigma$, the flow profiles are almost the same,
so that the IC emission processes, which does not directly depend on the
magnetic field, are common.
On the other hand, above $10^{23}$ Hz, the softening of the photon spectrum
with increasing  $\sigma$ is seen.
The softening is caused by the decline of the cut-off energy $E_{\rm cut}(r_{\rm N})$ due to synchrotron cooling,
though $E_{\rm max}$ is higher for a larger $\sigma$.
Above $\sigma=10^{-4}$, the short advection time leads
to the reduction of the IC flux.
The drop of the cooling efficiency
due to the short advection time causes the spectral hardening of the IC component.
While the flux decrease of the synchrotron component due to the reduction
of the advection time is mitigated by the magnetic field growth,
the IC component more rapidly falls with $\sigma$ than the synchrotron one.

In the right panel of Figure \ref{fig:sigma_dependence},
the X-ray surface brightness profile is shown.
The size of the X-ray nebula contracts with increasing $\sigma$.
The dependence of $\sigma^{-1/6}$
obtained in the previous subsection
is a reasonable approximation.

\section{Application to Observed Sources}\label{sec:application}
Here, we apply our model to the observed sources, 3C 58 and G21.5-0.9,
for which significant data sets are available to constrain the model parameters.
Moreover, in both the two PWNe, the extent of the X-ray emission is close to the radio.
We argue both the entire spectra and spatial profiles for those objects.

The images of 3C 58 were obtained in the radio wavelengths \citep[e.g.,][]{1988ApJ...327..845R}
and X-ray band \citep[e.g][]{2004ApJ...616..403S}.
The radial profile of photon index in the X-ray band was also obtained.
The extents of the radio and X-ray images are similar as $\sim 5'\times 9'$.
From the distance $D \simeq 2$ kpc \citep{2013AA...560A..18K}, we adopt $r_{\rm N}=2$ pc \citep{2013MNRAS.429.2945T}.
The spin period and its time derivative for the central pulsar of 3C 58 (PSR J0205+6449)
are $65.7\ {\rm ms}$ \citep{2002ApJ...568..226M, 2002ApJ...571L..41C} and $1.94\times10^{-13}\ {\rm s\ s^{-1}}$ \citep{liv09}, respectively.
The spin-down luminosity is estimated as $2.7\times10^{37}\ {\rm erg\ s^{-1}}$,
assuming $10^{45}\ {\rm g\ cm^2}$ for the moment inertia of the pulsar.

G21.5-0.9 shows spherical structures in the radio \citep{2008MNRAS.386.1411B}
and X-ray \citep{2005AdSpR..35.1099M,2010ApJ...724..572M,2006ApJ...637..456C} images.
The radio and X-ray sizes of the PWN ($\sim 40''$ in radius) are almost the same again.
Adopting the distance $D=4.8\ {\rm kpc}$ \citep{2008MNRAS.391L..54T}, we have $r_{\rm N}=0.9$ pc.
PSR J1833-1034, the central object of G21.5-0.9, has a spin period $61.9$ ms \citep{Gup05} and its derivative
$2.02\times10^{-13}\ {\rm s\ s^{-1}}$ \citep{Roy12}, from which we obtain
the spin-down luminosity $L_{\rm sd}=3.5\times10^{37}\ {\rm erg\ s^{-1}}$.

\begin{figure*}[!htbp]
	\begin{center}
		$\begin{array}{cc}
		\includegraphics[width=0.5\textwidth]{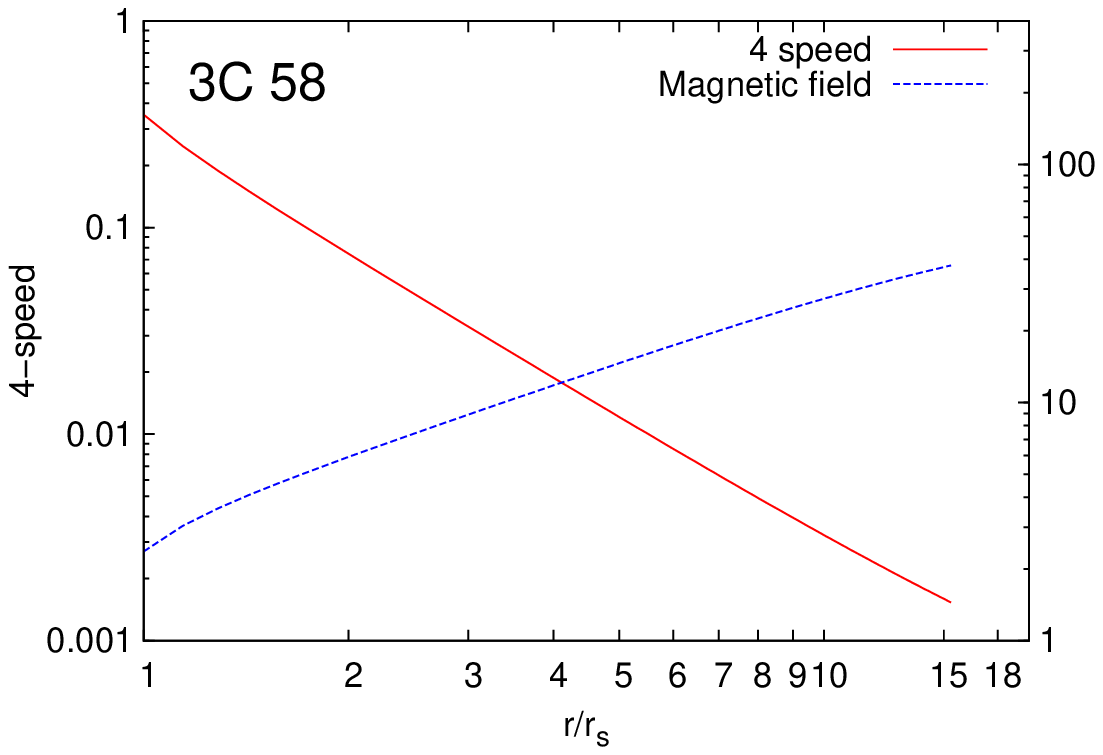} &
		\includegraphics[width=0.5\textwidth]{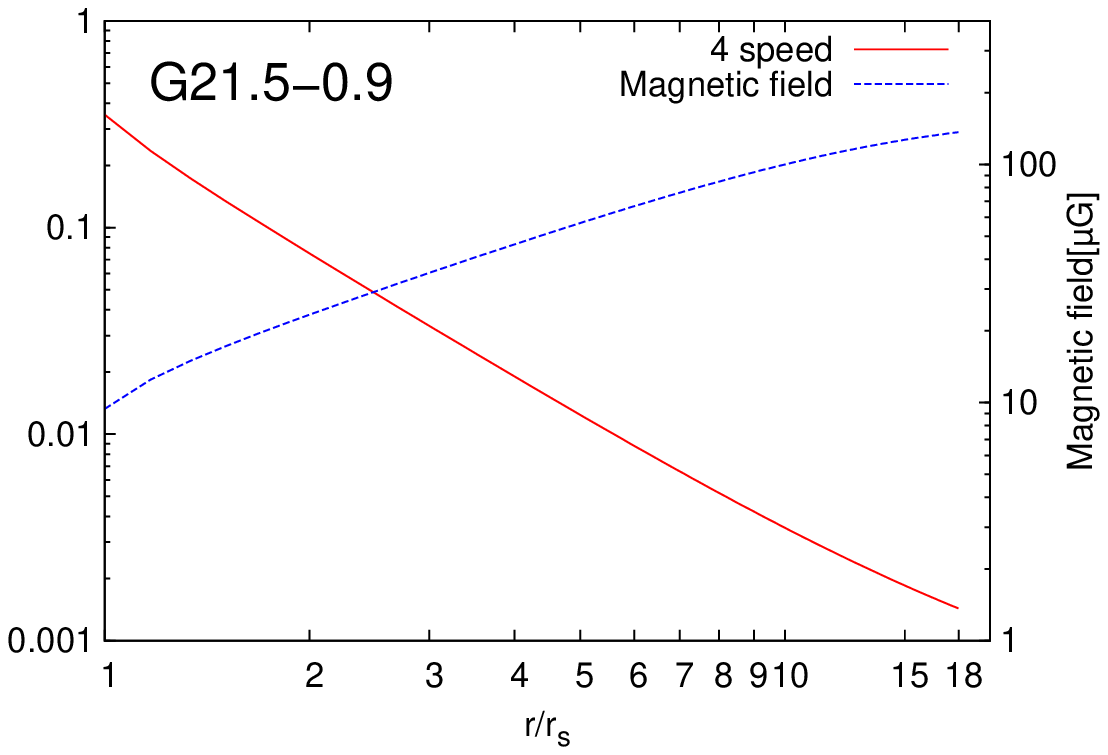}
		\end{array}$
	\end{center}
	\caption{Radial profiles of the four-speed $u(r)$ (left axis) and the magnetic field $B(r)$ (right axis)
in the broadband model (see Table \ref{BestFitParameter})
for 3C 58 (left panel) and G21.5-0.9 (right panel).}
	\label{fig:fit:uB}
\end{figure*}

\begin{figure*}[!htbp]
	\begin{center}
		$\begin{array}{cc}
		\includegraphics[width=0.5\textwidth]{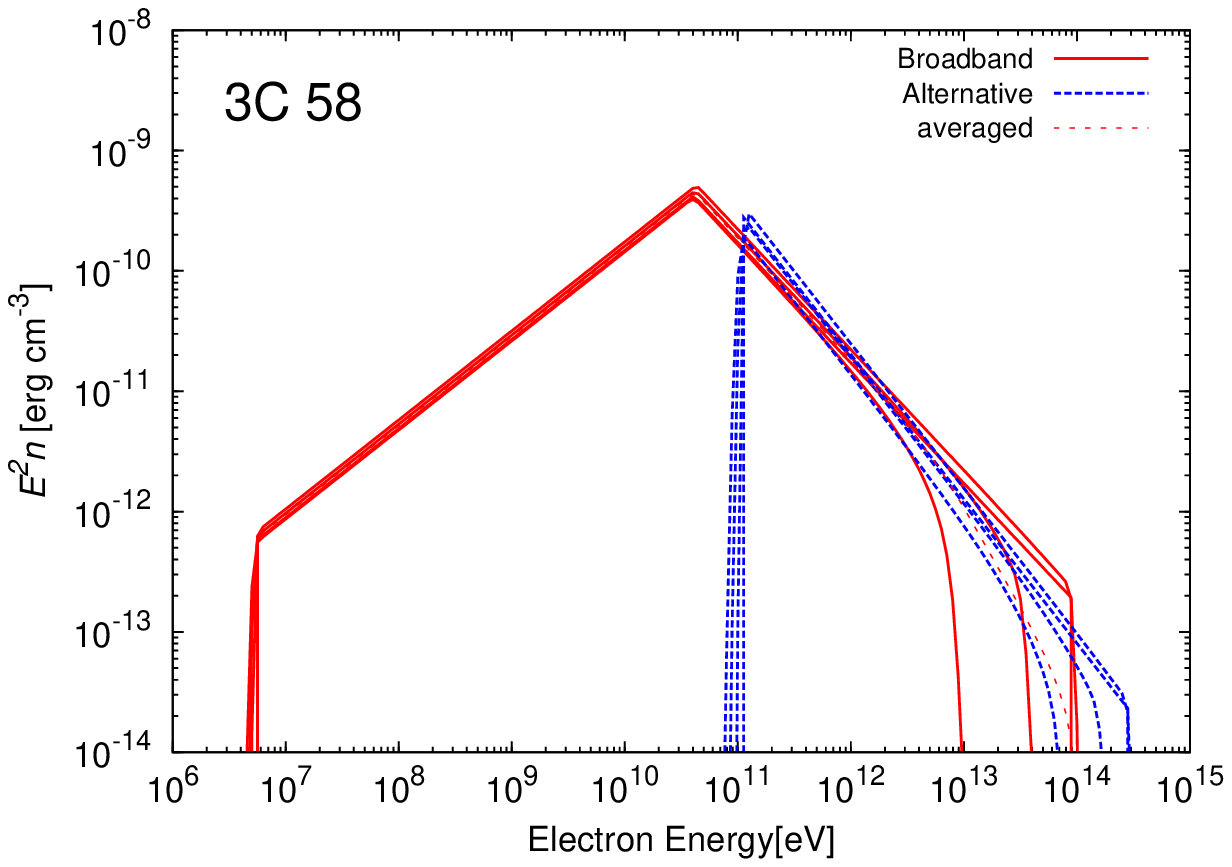} &
		\includegraphics[width=0.5\textwidth]{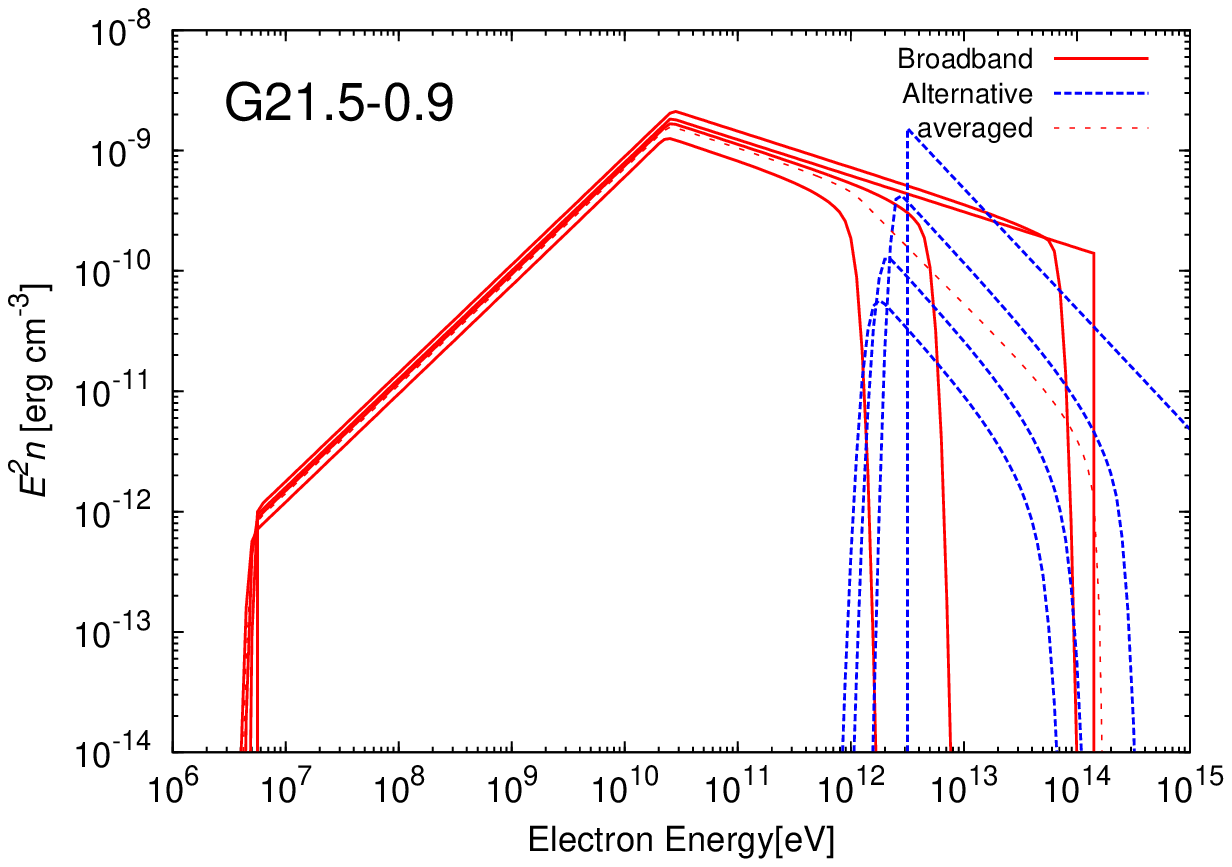}
		\end{array}$
	\end{center}
	\caption{Evolutions of the particle spectra for 3C 58 (left panel) and G21.5-0.9 (right panel)
in the broadband (red solid) and alternative (blue dashed) models.
The different lines correspond to $r=r_{\rm s}$, $5 r_{\rm s}$, $10 r_{\rm s}$, $r_{\rm N}\simeq 15 r_{\rm s}$ for 3C 58,
and $r=r_{\rm s}$, $6 r_{\rm s}$, $12 r_{\rm s}$, $r_{\rm N}= 18 r_{\rm s}$ for G21.5-0.9
(right to left).
The thin dashed lines represent the volume-averaged spectra in the broadband models.}
	\label{fig:fit:electron}
\end{figure*}

The parameters to fit the spectra for the two PWNe are summarized in Table \ref{BestFitParameter}.
See Figure \ref{fig:soft_photon} for the ISRFs taken from GALPROP v54.1 for the two cases.
First, we discuss the parameter sets denoted with ``broadband'' in the Table \ref{BestFitParameter} (hereafter we call them broadband model).
The resultant radial profiles of $u(r)$ and $B(r)$ are shown in Figure \ref{fig:fit:uB}.
In both the two PWNe, $r_{\rm eq}$ is outside the nebula radius $r_{\rm N}$ in our parameter sets.
The particle spectra in Figure \ref{fig:fit:electron} show the evolution of $E_{\rm cut}$
as discussed in Section \ref{Characteristics}.
The volume averaged spectra (dashed lines) $\bar{n}(E)$ are well expressed by broken power-laws.
The differences of the particle spectral index above the cooling break energy are 0.67 for both the cases.
Given the particle energy $E$, Equation (\ref{eq:cut}) implies the maximum radius 
those particles survive $r_E \propto E^{-1/5}$.
The differences in the index seem consistent with a naive estimate $\bar{n}(E) \propto E^{-p_2} r_E^3$.

\begin{table*}[hbtp]
	\begin{center}
		\begin{tabular}{lcrrrr}
			\hline\hline
			 &      & \multicolumn{2}{c}{3C 58} & \multicolumn{2}{c}{G21.5-0.9} \\
			Given parameters &   Symbol   & \multicolumn{1}{c}{broadband} & \multicolumn{1}{c}{alternative}
             & \multicolumn{1}{c}{broadband} & \multicolumn{1}{c}{alternative} \\
			\hline
			Spin-down luminosity {\rm (erg~s$^{-1}$)} & $L_{\rm sd}$ & \multicolumn{2}{c}{$3.0\times10^{37}$} &
              \multicolumn{2}{c}{$3.5\times10^{37}$} \\
			Distance (kpc) & $D$ & \multicolumn{2}{c}{$2.0^{\rm a}$} & \multicolumn{2}{c}{$4.8^{\rm b}$} \\
			Radius of the nebula (pc) & $r_{\rm N}$ & \multicolumn{2}{c}{$2.0$} & \multicolumn{2}{c}{$0.9$} \\
			\hline
			Fitting parameters\\
			\hline
			Break energy (eV) & $E_{\rm b}$ & $4.1\times10^{10}$ & $1.1\times10^{11}$
               & $2.6\times10^{10}$ & $3.1\times10^{12}$\\
			Low energy power-law index & $p_1$ & $1.26$ &   & $1.1$ &  \\
			High energy power-law index & $p_2$ & $3.0$ & $3.2$ & $2.3$ & $3.0$\\
			Radius of the termination shock (pc) & $r_{\rm s}$ & $0.13$ & $0.26$ & $0.05$ & $0.1$ \\
			Magnetization parameter & $\sigma$ & $1.0\times10^{-4}$ & $1.0\times 10^{-3}$ & $2.0\times 10^{-4}$ & $3.0\times 10^{-2}$\\
			\hline
			Obtained parameters\\
			\hline
            Initial bulk Lorentz factor & $\gamma_{\rm u}$ & $7.3\times10^3$ & $5.5\times10^{5}$ & $2.1\times10^{4}$ & $1.8\times10^{7}$\\
            Pre-shock density (cm$^{-3}$) & $n_{\rm u}$ & $1.1 \times 10^{-11}$ & $5.1\times10^{-16}$ & $1.1\times10^{-11}$ & $3.5\times10^{-18}$\\
			Pre-shock magnetic field ($\mu$G) & $B_{\rm u}$ & $0.79$ & $1.2$ & $3.1$ & $19$ \\
			Maximum Energy (eV) & $E_{\rm max}$ & $9.5\times10^{13}$ & $3.0\times10^{14}$ & $1.4\times10^{14}$ & $1.7\times10^{15}$ \\
			Advection time (yr) & $t_{\rm adv}$ & $1500$ & $330$ & $800$ & $38$\\
			Averaged magnetic field ($\mu$G) & $B_{\rm av}$ & $31$ & $21$ & $120$ & $61$ \\
			Total pressure at $r=r_{\rm N}$ (erg~cm$^{-3}$) & $p_{\rm tot}$ & $3.7\times10^{-10}$ & $8.3\times10^{-11}$ & $2.1\times10^{-9}$ & $1.2\times10^{-10}$\\
			$r_{\rm eq} / r_{\rm N}$ &  & $3.8$ & $2.4$ & $2.3$ & $0.37$\\
			\hline\hline
		\end{tabular}
	\end{center}
{\scriptsize 
$^{\rm a}$ \citet{2013AA...560A..18K}; $^{\rm b}$ \citet{2008MNRAS.391L..54T}.
}
	\caption{Parameters in our calculations.}
	\label{BestFitParameter}
\end{table*}

\begin{figure*}[!htbp]
	\begin{center}
		$\begin{array}{cc}
		\includegraphics[width=0.5\textwidth]{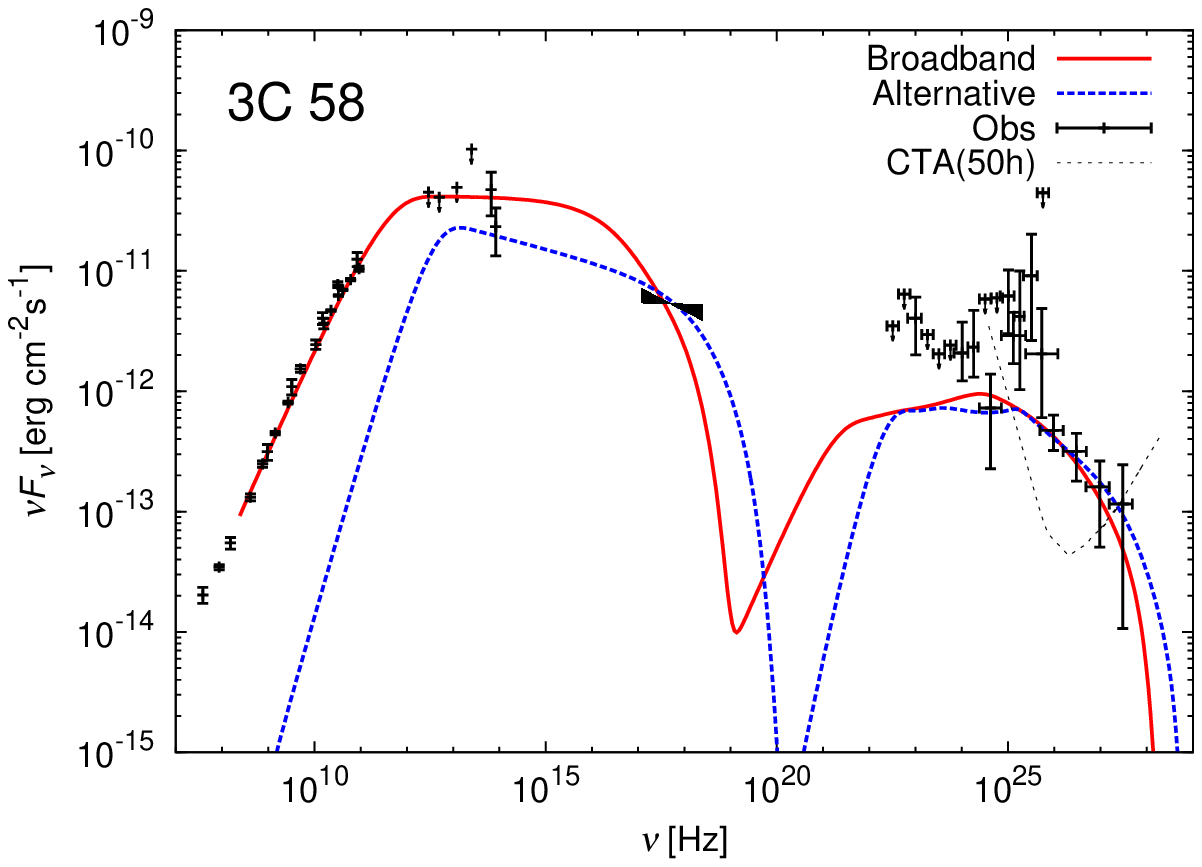} &
		\includegraphics[width=0.5\textwidth]{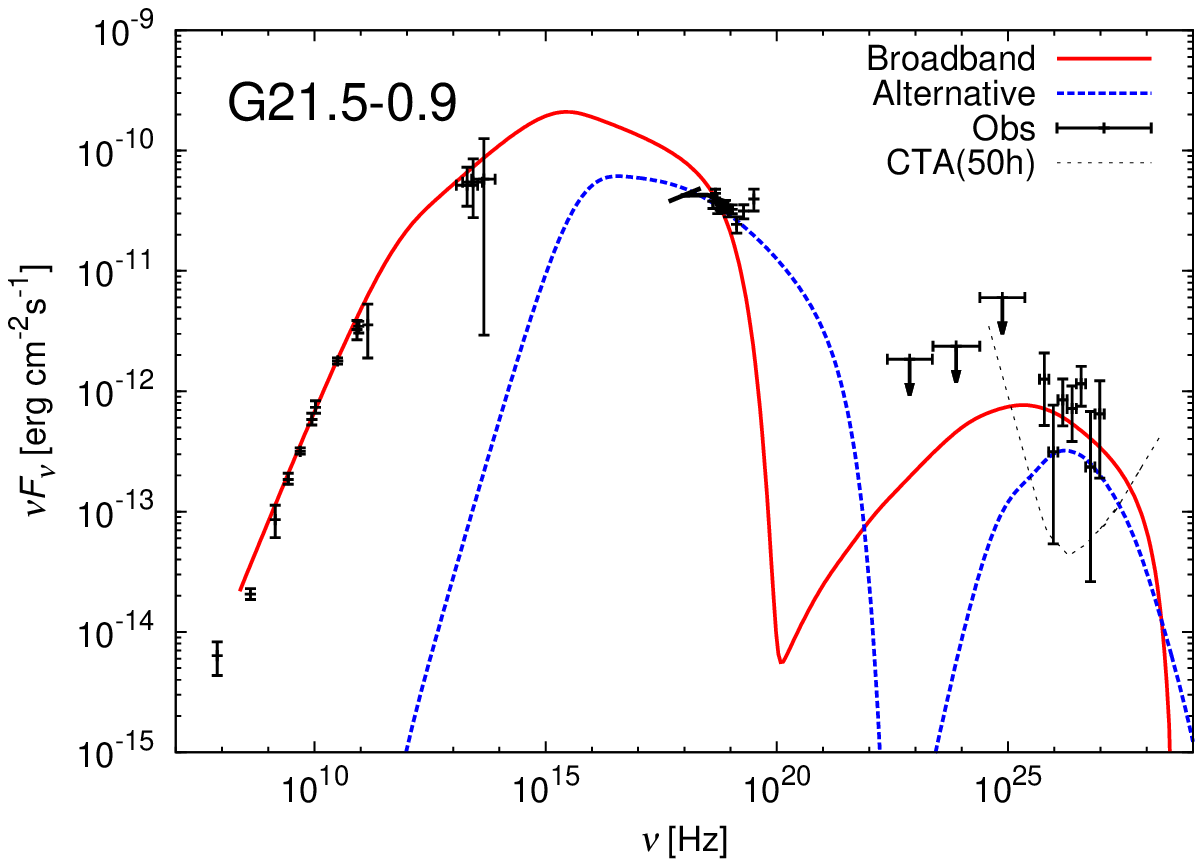}
		\end{array}$
	\end{center}
	\caption{Entire photon spectra for 3C 58 (left panel) and G21.5-0.9 (right panel).
The data points are taken from \citet{2011ApJS..192...19W} (radio),
\citet{1994ApJS...90..817G}, \citet{2008ApJ...676L..33S} (IR), \citet{2000PASJ...52..875T} (X),
\citet{2013ApJS..208...17A} (GeV),
and \citet{2014AA...567L...8A} (TeV) for 3C 58,
and \citet[][]{1989ApJ...338..171S} (Radio), \citet{1998MmSAI..69..963G} (IR),
\citet{2011AA...525A..25T}, \citet{2014ApJ...789...72N}, \citet{2009MNRAS.393..527D} (X),
\citet{2011ApJ...726...35A} (GeV), and \citet{2008ICRC....2..823D} (TeV) for G21.5-0.9.
While the red solid lines represent the broadband models,
the blue dashed lines represent the alternative models (see text), in which the radio/IR data are disregarded.}
	\label{fig:fit:spectrum}
\end{figure*}

Figure \ref{fig:fit:spectrum} shows the volume-integrated photon spectra for the two PWNe.
Our models roughly reproduce the entire structures of the spectra.
In 3C 58, the data points obtained with {\it Fermi} \citep{2013ApJS..208...17A}
may contain large systematic errors due to the emission from the central pulsar
so that we treat those data as upper limits.
The model spectrum in the X-ray range is apparently softer than the X-ray data in 3C 58 due to the cooling effect.
As discussed in Section \ref{dep:sig}, $\nu_{\rm c}$
can be higher
than the X-ray energy range by adopting a larger $\sigma$ or conversely lower $\sigma$.
In such cases, the X-ray model spectrum may be as hard as observed one.
However, when we adopt a lower $\sigma$ to make $\nu_{\rm c}$ above the X-ray frequency,
the synchrotron component does not extend to the X-ray energy
as Equation (\ref{eq:nucut}) indicates.
We also do not find a consistent high-$\nu_{\rm c}$ model with a very large $\sigma$ or a slightly large $r_{\rm s}$,
for which the radio and X-ray fluxes are hard to be reproduced simultaneously.
One may suppose that a smaller $p_2$ can agree with the observed X-ray spectral index,
even if $\nu_{\rm c}$ is below the X-ray frequency.
The extrapolation from the X-ray data requires $\nu_{\rm c}<10^{14}$ Hz to
make $\nu_{\rm b}$ above the radio data points.
Such a low $\nu_{\rm c}$ is hard to be realized in this model
(see the $\nu_{\rm c}$-turnover in Figure \ref{fig:sigma_dependence});
the radio or X-ray flux becomes inconsistent for such extreme parameter sets.
Therefore, our model spectra
cannot be reconciled with the X-ray spectral index.

A similar problem to the case in 3C 58 arises in the X-ray spectrum
of G21.5-0.9.
When we fit a model with $p_2=2$, which yields
a flat spectrum ($\nu L_\nu \propto \nu^0$) above $\nu_{\rm c}$,
a much lower $\sigma$ is required to adjust the X-ray flux.
For such a low $\sigma$, $\nu_{\rm cut}$ becomes lower than the X-ray band.

\begin{figure*}[!htbp]
	\begin{center}
		$\begin{array}{cc}
		\includegraphics[width=0.5\textwidth]{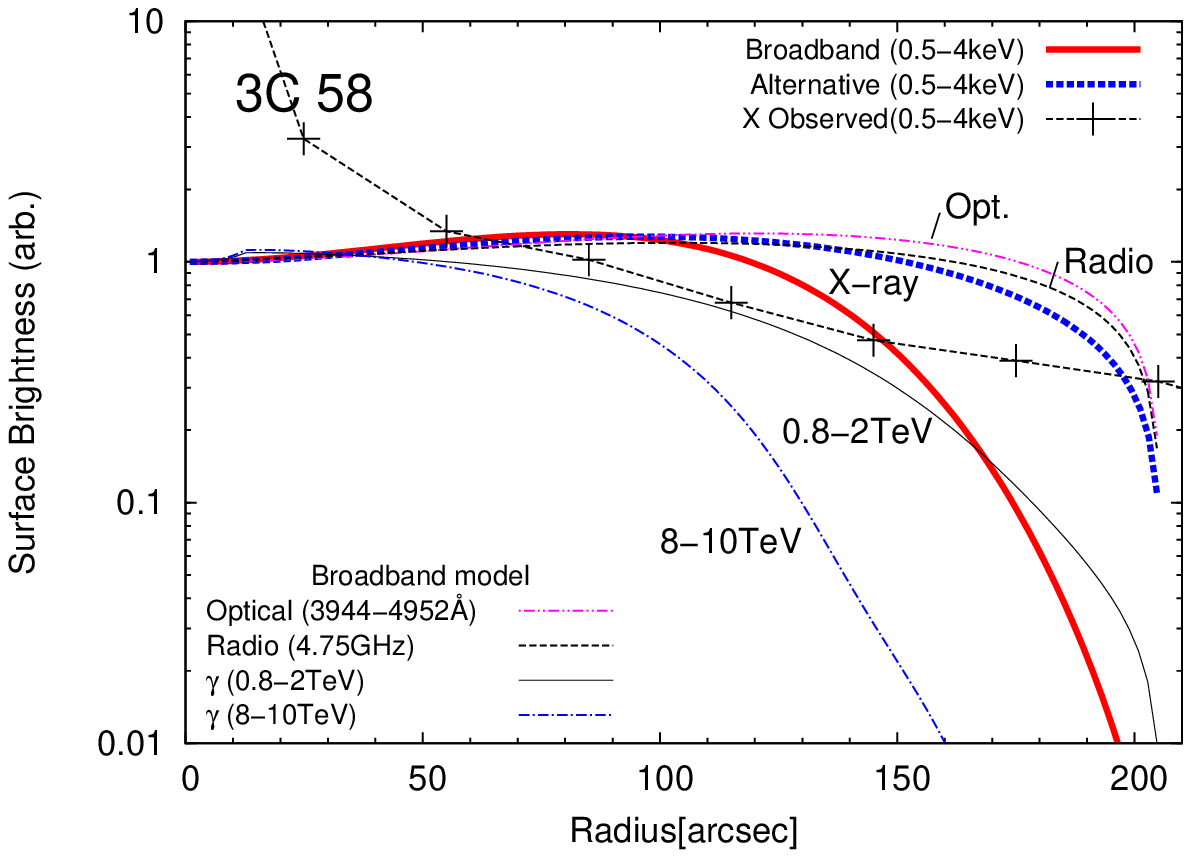} &
		\includegraphics[width=0.5\textwidth]{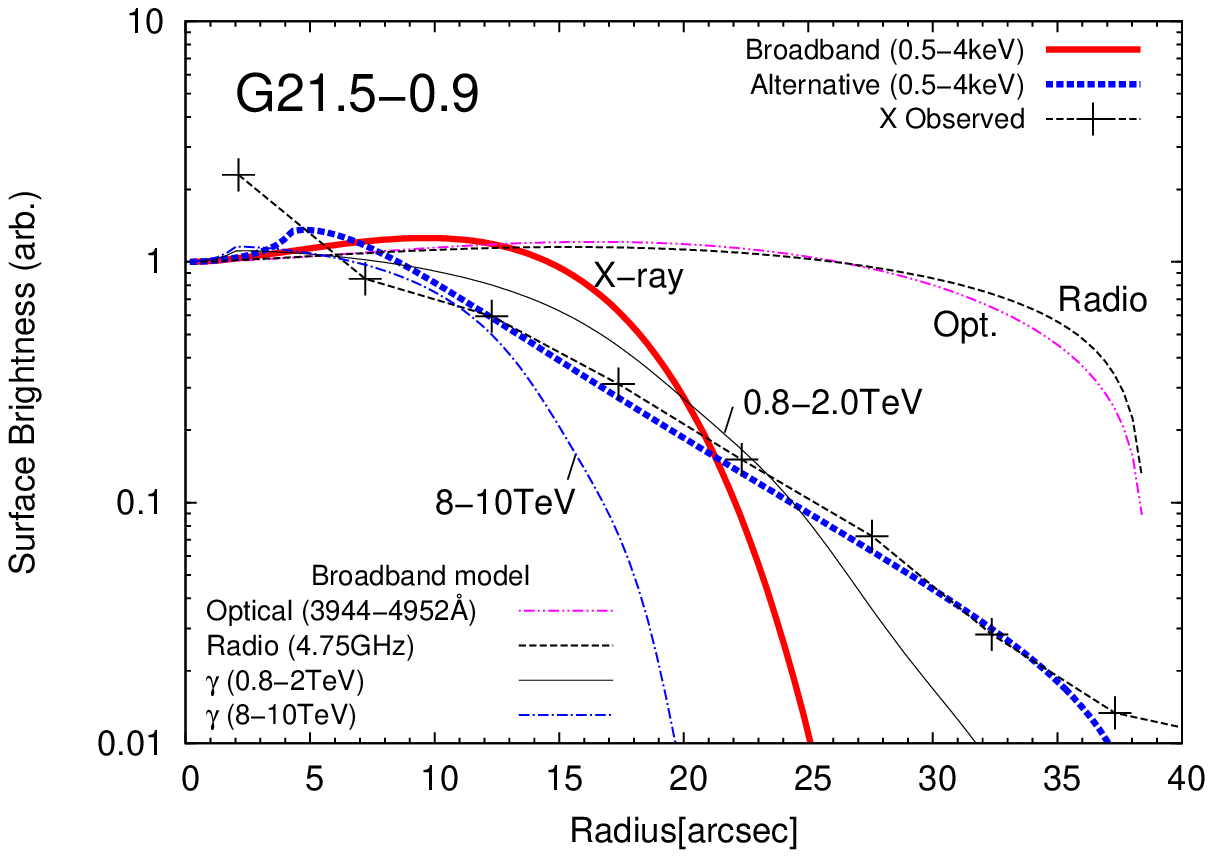}
		\end{array}$
	\end{center}
	\caption{
Radial profiles of the surface brightness in various frequencies
for 3C 58 (left panel) and G21.5-0.9 (right panel).
The X-ray data points are taken from \citet{2004ApJ...616..403S}
and \citet{2005AdSpR..35.1099M} for 3C 58 and G21.5-0.9, respectively.
The thick lines are the X-ray surface brightnesses for the broadband (red solid) and
alternative (blue dashed) models.}
	\label{fig:fit:brightness}
\end{figure*}

\begin{figure*}[!htbp]
	\begin{center}
		$\begin{array}{cc}
		\includegraphics[width=0.5\textwidth]{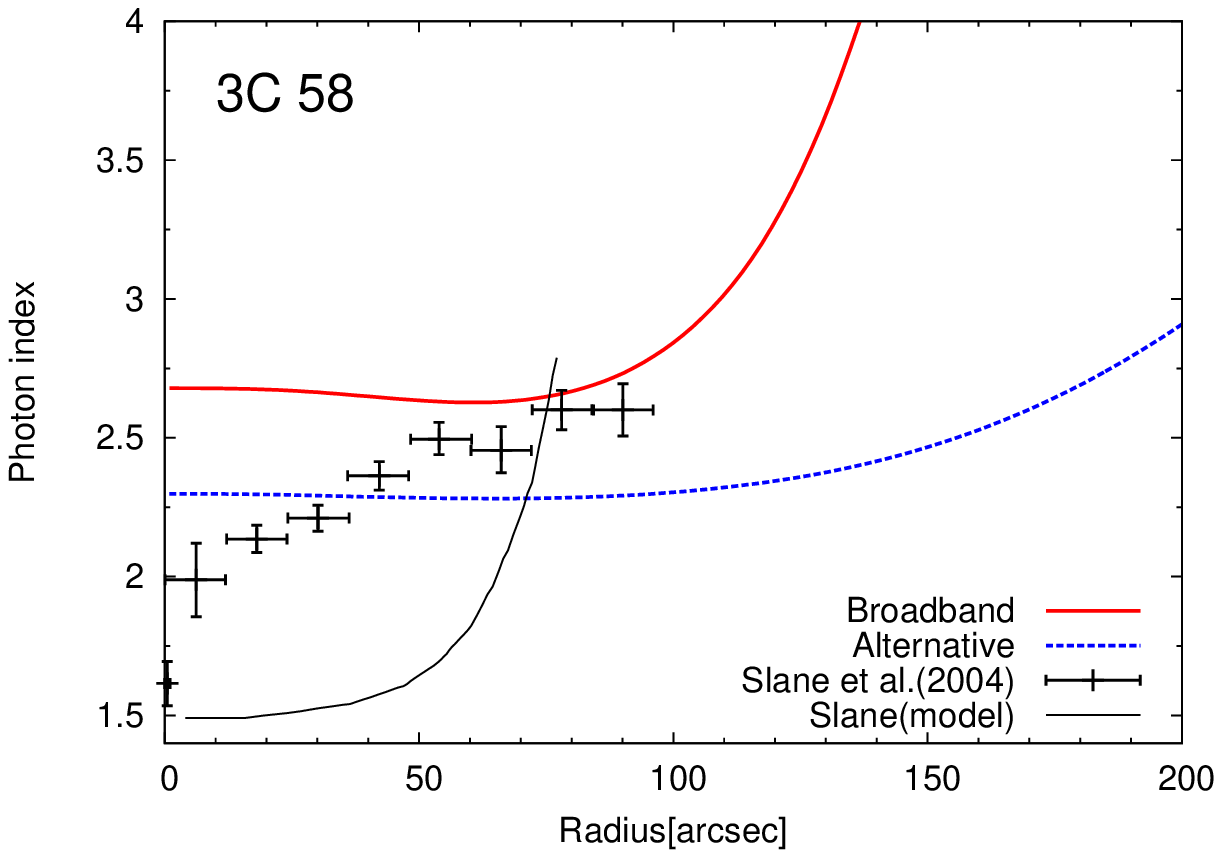} &
		\includegraphics[width=0.5\textwidth]{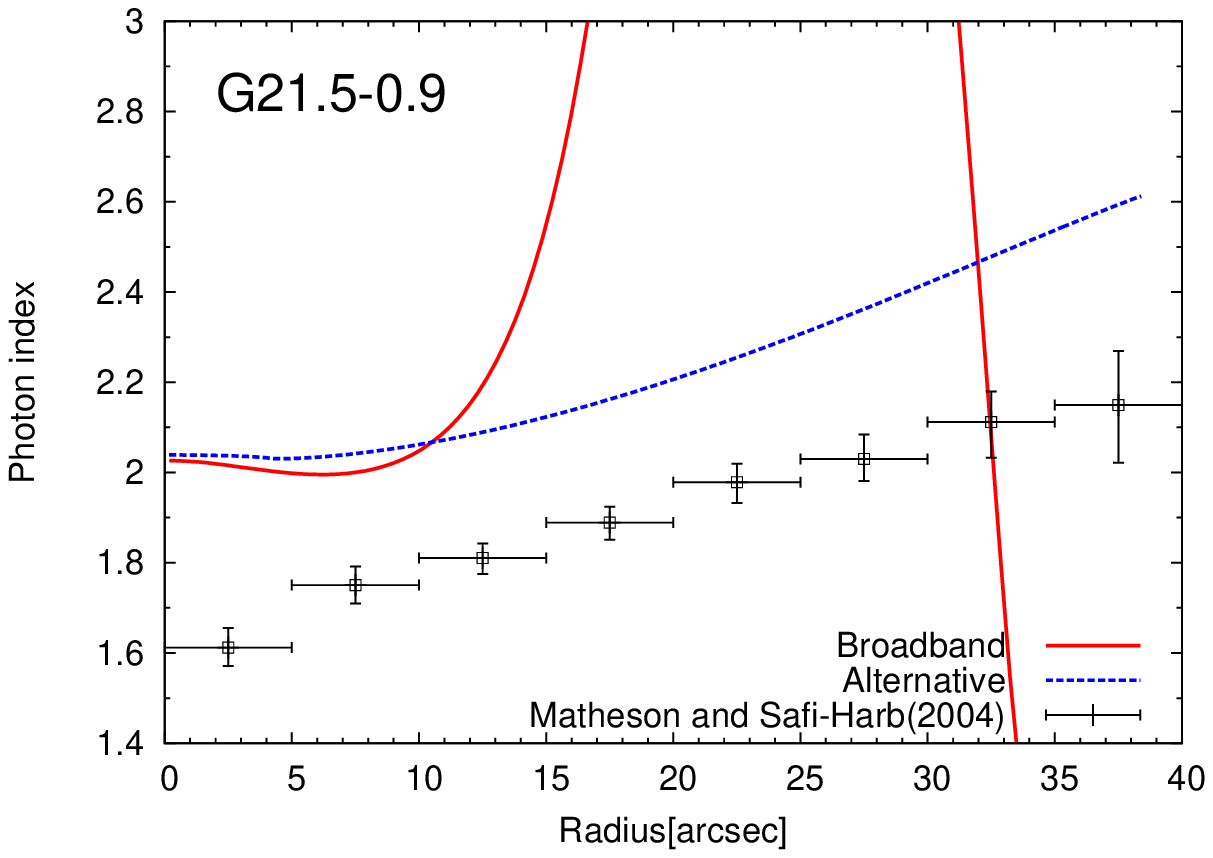}
		\end{array}$
	\end{center}
	\caption{Radial profiles of photon indices in 0.5-10.0 keV range for 3C 58 (left panel) and G21.5-0.9 (right panel)
in the broadband (red solid) and alternative (blue dashed) models.
The model and data points in \citet{2004ApJ...616..403S} are also plotted for 3C 58.
The data points for G21.5-0.9 are taken from \citet{2005AdSpR..35.1099M}.}
	\label{fig:fit:index}
\end{figure*}

In those two objects, we are forced to have $\nu_{\rm c}$ below the X-ray band.
As a result, the X-ray spectra show softer shapes than the observed ones.
The X-ray extents are more compact than the radio images (Figure \ref{fig:fit:brightness}).
The radial profiles of photon indices in 0.5-10.0 keV range (Figure \ref{fig:fit:index}) also deviate
from the observed data.
However, the discrepancy in 3C 58 is not so prominent compared to the model curve
by \citet{2004ApJ...616..403S} based on \citet{Rey03}.
Note that the radial profiles of photon indices in optical (3944-4952$\AA$) and radio (4.75 GHz) band for two objects
do not depend on angular distance from the pulsar.
Since the $\nu_{\rm c}$ is higher than the frequencies of these bands in the parameters of the broadband models,
the pairs can emit radio and optical photons without the cooling effect all over the nebula,
thus the emission has same spectral index in each radial position.

The advection time of G21.5-0.9 well agrees with the age $870$ yr \citep{2008MNRAS.386.1411B}.
However, \citet{2006ChJAA...6..625W} argued that this object associates with BC48 guest star and its age is thus about $2000$ yr.
In this case, $t_{\rm adv}$ in the broadband model becomes less than half of the age.
In 3C 58, if this object associates with SN 1181 \citep{1971QJRAS..12...10S},
$t_{\rm adv}$ is about 2 times larger than the age.
The characteristic ages of these objects are $5370$ yr for 3C 58 and $4850$ yr for G21.5-0.9, respectively.
Notify that the characteristic age tends to be longer than the actual pulsar age,
especially for young pulsars.
Meanwhile, the previous one-zone time dependent models have obtained the ages.
For 3C 58, \citet{buc11}, \citet{tor13} and \citet{2013MNRAS.429.2945T} obtained $t_{\rm age} \sim 2000$ yr, 
which is comparable with $t_{\rm adv}$ in our broadband model.
However, \citet{buc11} and \citet{Tor14} adopted the different value of the distance to the object,
and these three studies did not include the data of \citet{2014AA...567L...8A}.
A direct comparison of the age with our advection times does not seem meaningful so much.
On the other hand, for G21.5-0.9, $t_{\rm age}$ was estimated to be $870$ yr \citep{vor13,Tor14} or 1000yr \citep{2011ApJ...741...40T} in one-zone models.
Those are close to our estimate.
These studies adopted the same condition for the distance and the observed flux,
so that the coincidence of the age and $t_{\rm adv}$ encourages our 1-D model.

Next, we discuss the case, where the radio/IR/optical emission can be treated separately as an additional component.
In the broken power-law spectrum for the pair injection,
the low-energy portion dominates the number.
In the case of the Crab Nebula, the required particle number to reconcile the radio flux
is much larger than the theoretically expected value \citep{2010ApJ...715.1248T,2011ApJ...741...40T}.
\citet{1996MNRAS.278..525A} and \citet{2015MNRAS.449.3149O} treated the low-energy component
as a different component from the wind particles in their calculations.
Thus, as an alternative model, 
we assume that the low-energy particle component responsible for the radio/IR emission has a different origin
from the high-energy component.
In the alternative models,
we incorporate only the high energy particles above $E_{\rm b}$ neglecting the emission below the optical band..

Blue dashed lines in Figure \ref{fig:fit:spectrum}-\ref{fig:fit:index} are for the alternative models,
whose model parameters are summarized in Table \ref{BestFitParameter}.
We adopt a slightly large $r_{\rm s}$ and large $\sigma$, which lead to X-ray extents
consistent with observation as shown in Figure \ref{fig:fit:brightness}.
The difference of the X-ray profiles in the two PWNe is attributed to
the effect of the adiabatic cooling in G21.5-0.9 as shown in Figure \ref{fig:fit:electron}.
Since we have adopted a larger $\sigma$ enough to establish $ r_{\rm eq} < r_{\rm N} $ for G21.5-0.9,
a signature of adiabatic cooling appears.
Although those models seem to reproduce the observed X-ray surface brightness and X-$\gamma$ fluxes,
the resultant advection times become very short (see Table \ref{BestFitParameter}).

	\section{Discussion}\label{sec:Discussion}
As shown in Section \ref{sec:application}, 
we have fitted the entire spectra of 3C 58 and G21.5-0.9.
The obtained $\sigma$ by fitting the entire spectrum of nebula in the broadband models is about 10 times smaller than the conceivable value in the Crab Nebula
and obtained $r_{\rm s}$ is similar to the value of the Crab Nebula.
Our 1-D model has difficulty to reproduce both of the hard spectra and the size of PWNe in X-rays.
As discussed in Section \ref{sec:application}, our model is hard to avoid the spectral softening due to the cooling effect
in X-ray range.
As a result, the X-ray nebula size becomes more compact than the observed extents.
The 1-D model should be improved by introduction of possible physical processes, such as the spatial diffusion of
high energy particles, reacceleration by turbulences, and amplification/dissipation of the magnetic field.

The 1-D model tends to lead a lower $\sigma$ than the values derived from one-zone models.
One-zone time-dependent models
resulted in $\sigma \sim 0.03$--$0.5$ \citep{buc11,2013MNRAS.429.2945T,tor13}
and $0.01$--$0.2$ \citep[][]{2011ApJ...741...40T,vor13,Tor14}
for 3C 58 and G21.5-0.9, respectively,
while the value in our paper is $\sigma \sim 10^{-4}$.
Since the magnetic field increases with radius in the 1-D model,
the average magnetic field $B_{\rm av}$ is consistent with the previous one-zone models
(see also Equations (\ref{eq:Ec_ave}) and (\ref{eq:nuc_ave})).
On the other hand,
$\nu_{\rm cut}$ is determined by the magnetic field near the shock rather than $B_{\rm av}$,
in contrast to the cooling break $\nu_{\rm c}$.
Therefore, the one-zone model overestimates the maximum synchrotron frequency about
$B_{\rm av}/B_{\rm d}$ times higher than the 1-D model.
Note that hard X-ray observations around $\nu_{\rm cut}$ are also interesting to investigate 
how the maximum energy of non-thermal pairs is determined.
Although $E_{\rm max}$ is constrained by the size of $r_{\rm s}$
in our case as shown in Equation (\ref{eq:ene_max}),
$\nu_{\rm cut} \sim 100$ MeV in the Crab Nebula
implies that $E_{\rm max}$ is determined by the balance of the
acceleration and cooling times \citep[e.g.,][]{1996ApJ...457..253D}.

The fitted shock radius $r_{\rm s}$ of 3C 58 is twice as large as that of G21.5-0.9.
The total pressures at $r=r_{\rm N}$ in our models are
$p_{\rm tot, 3C~58}\sim3.7\times10^{-10}$ erg~cm$^{-3}$ and $p_{\rm tot, G21.5}\sim 2.1\times10^{-9}$ erg~cm$^{-3}$.
From the pressure balance, the large $p_{\rm tot}$ implies the large plasma pressure of the surrounding remnants of their supernovae.
\citet{2004ApJ...616..403S} obtained $kT_e \sim 0.23$ keV and $n_{\rm SNR} \sim 0.38$ cc$^{-1}$ for 3C 58 and then its pressure $\sim1.4 \times 10^{-10}$ erg~cm$^{-3}$.
\citet{2010ApJ...724..572M} obtained $kT_e\sim 0.3$ keV and $n_{\rm SNR}\sim0.63$ cm$^{-3}$ for G21.5-0.9 and then its pressure $\sim3.0 \times 10^{-10}$ erg~cm$^{-3}$.
While the pressure values may be not so robust, this indicates that the surrounding SNR of G21.5-0.9 has higher pressure than 3C 58.
In addition, the fact that the bright shell-like SNR is clearly seen in G21.5-0.9
also supports that the pressure for G21.5-0.9 would be higher than that for 3C 58.

In the broadband models, although we reproduce the flux levels of the entire spectrum, the X-ray spectral indices disagree with the observations.
We argue the alternative models,
in which the emission in radio and optical is assumed to be different from the direct emission from the pulsar wind \citep{1996MNRAS.278..525A,2014MNRAS.438.1518O,2015MNRAS.449.3149O}.
As shown in Table \ref{BestFitParameter},
the obtained $\sigma$ in the alternative models tends to be larger than the value in the broadband models.
This tendency is similar to some one-zone models \citep{tor13, vor13}.
The time-dependent model of \citet{tor13} introduced an order of magnitude larger energy density of the ISRF (i.e., the lager magnetic field strength) than ours
in order to reproduce the X-ray spectral index of 3C 58.
In the model of \citet{vor13}, the X-ray spectrum of G21.5-0.9
was also reproduced by a strong magnetic field ($230\mu$G),
and hard spectral index ($p_2=2.0$).
Note that their predicted GeV flux seems above the {\it Fermi}
upper-limit \citep{2011ApJ...726...35A}.
The cooling break was set well below keV range in those models,
while it is hard to set $\nu_{\rm c}$ low enough in our model,
because the high $\sigma$ leads to a shorter advection time than the cooling time scale
of low-energy particles as discussed in Section \ref{dep:sig}.
The temporal evolution of the magnetic field in one-zone models causes the gradual hardening of the particle spectrum
\citep[see also][]{2010ApJ...715.1248T},
which is favorable to fit the X-ray data differently from our steady model.

Since a larger $\sigma$ is required in the alternative models,
the resultant short advection time prevents high-energy particles from cooling before reaching the edge of the nebula.
However, such a short advection time may contradict the age of the PWNe.
To validate $t_{\rm age}\gg t_{\rm adv}$, the efficient particle escape
at the nebula surface should be required.
Although the large amount of the escaped high-energy particles should emit photons outside the PWNe, such a signature outside PWNe has not been claimed.
For example, the model of \citet{hol12B}, in which the radial velocity profile is artificially tuned,
also implies $\sim 100$ yr for the advection time in G0.9+0.1,
though the age is more than kyr.
We should carefully note the advection time in modeling the outflow property (see Equation (\ref{eq:def_tadv})).
Even for the models including the effect of the spatial diffusion \citep{tan12,por16},
a short diffusion time scale may be required to reproduce the X-ray surface brightness.
This leads to the same problem in the short advection time case.
If the outer supernova ejecta efficiently confines the PWN,
the fast outflow implied in the high $\sigma$ model should be decelerated near the edge of the PWN,
and should induce turbulence inside the PWN.
As a result, the wind material may be efficiently mixed inside the PWN \citep[e.g.,][]{2014MNRAS.438..278P}.
In this case, the one-zone approximation may be rather adequate.

As shown in Figure \ref{fig:fit:brightness}, we calculated the radial radio profile in the broadband model.
They appear to be almost uniform profile in radius, and seem to agree with the observational facts \citep[e.g.,][]{2008MNRAS.386.1411B,2013MNRAS.431.2590B}.
Additionally, the radiation in $\nu < \nu_{\rm c}$ ($\sim 10^{15}$ Hz) shows the similar profile as radio one.
This is because the radial evolution of the density is common for the non-cooled particles.
If a clear distinction of surface brightness between the radio and optical is detected,
this would be the strong basis
of the hypothesis that the radio emission of PWNe has a different origin from the optical and X-rays.
We also calculate the radial profile of surface brightness in 0.8--2.0 TeV and 8--10 TeV.
The extents of 3C 58 and G21.5-0.9 in 0.8--2 TeV are larger than X-rays in the broadband models,
because the photon of $\sim 1$ TeV is emitted by non-thermal pairs with lower energies
than the energy of particles emitting the synchrotron radiation at $\nu\sim\nu_{\rm cut}$ \citep[c.f.,][]{2009ASSL..357..451D}.
Since the extent of 3C 58 is about $200''$,
CTA will be able to present the spatially resolved $\gamma$-ray map of 3C 58 in $\sim 1$ TeV.

\section{Conclusion}\label{sec:Conclusion}

\highlight{
We have revisited the 1-D steady model, and applied to the pulsar wind nebulae,
in order to find a parameter set consistent with both the entire photon spectrum and surface brightness profile.
It is still controversial whether the simple 1-D model reproduces observed properties of the PWNe other than the Crab Nebula or not.
As we have shown in Section \ref{sec:dependence}, both the entire photon spectrum and surface brightness profile
largely depend on the parameters, the uncertain shock radius $r_{\rm s}$ and the magnetization parameter $\sigma$.
The flux of inverse Compton component becomes dim with increasing $\sigma$.
In contrast, the synchrotron component is not a monotonic function of $\sigma$.
For the dependence on $r_{\rm s}$, while the synchrotron component becomes dim with increasing with $r_{\rm s}$, the IC component shows complicated behaviors.  
The X-ray size of a PWN becomes large with increasing $r_{\rm s}$ and decreasing $\sigma$.
}

\highlight{
We have fitted the entire spectrum of two observed sources 3C 58 and G21.5-0.9.
Calculating the radial profile of the surface brightness for those models,
we show that the resultant X-ray extents are significantly smaller than the observed sizes.
Furthermore, we have performed another parameter set called ``alternative'' model,
where we treat the radio and optical emissions as extra components.
The alternative models successfully reproduce the observed X-ray surface brightness and the X-ray and $\gamma$-ray fluxes.
However, those models imply too short advection time.
In summary, the 1-D model constructed by KC84s has severe difficulty to reproduce both the spectrum and spatial emission profile of PWNe consistently.
The model should be improved by taking some possible physical processes into consideration,
such as spatial diffusion of non-thermal particles, reacceleration by turbulences.
}

\section*{Acknowledgments}
First, we appreciate the anonymous referee for valuable comments.
We are grateful to Kento Sasaki for providing the calculation code.
This work is supported by Grants-in-Aid for Scientific
Research Nos. 15K05069 (TT and KA), 25400227, 16K05291 (KA), and 2510447 (SJT) from the Ministry
of Education, Culture, Sports, Science and Technology
(MEXT) of Japan.

\appendix\label{sec:appendix}

In order to check our model, we calculate the entire spectrum and the X-ray profile of the Crab Nebula and compare the result with \citet{1996MNRAS.278..525A} (hereafter AA96).
We treat the SSC approximately by assuming that the synchrotron photon exist in the nebula homogeneously.
The other differences of our model from AA96 are as follows.
In our model, the low-energy particles responsible for the radio emission are supplied from the pulsar wind,
while AA96 considered them as another component.
In AA96, the maximum energy of pairs was treated as a parameter of the model (but see Equation (\ref{eq:ene_max}) for our model).
Finally, AA96 introduced the correction factor $\kappa$, which is a parameter to adjust the ratio of the synchrotron flux to the SSC flux.
The parameter $\kappa$ (cf. AA96 adopted $\kappa\sim0.5$) may represent the effects of deviation from the spherical symmetry or inhomogeneity inside the PWN.

\begin{figure*}[!htbp]
	\begin{center}
		$\begin{array}{cc}
		\includegraphics[width=0.5\textwidth]{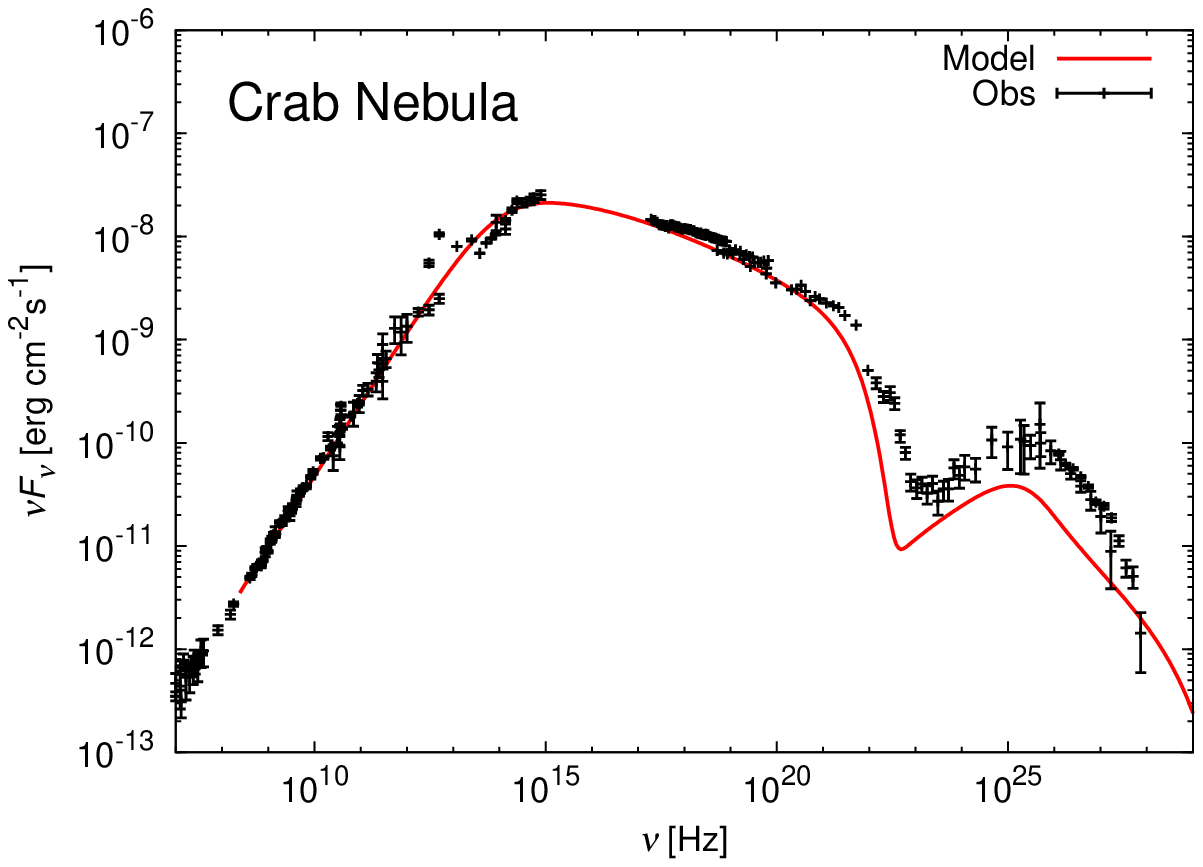} &
		\includegraphics[width=0.5\textwidth]{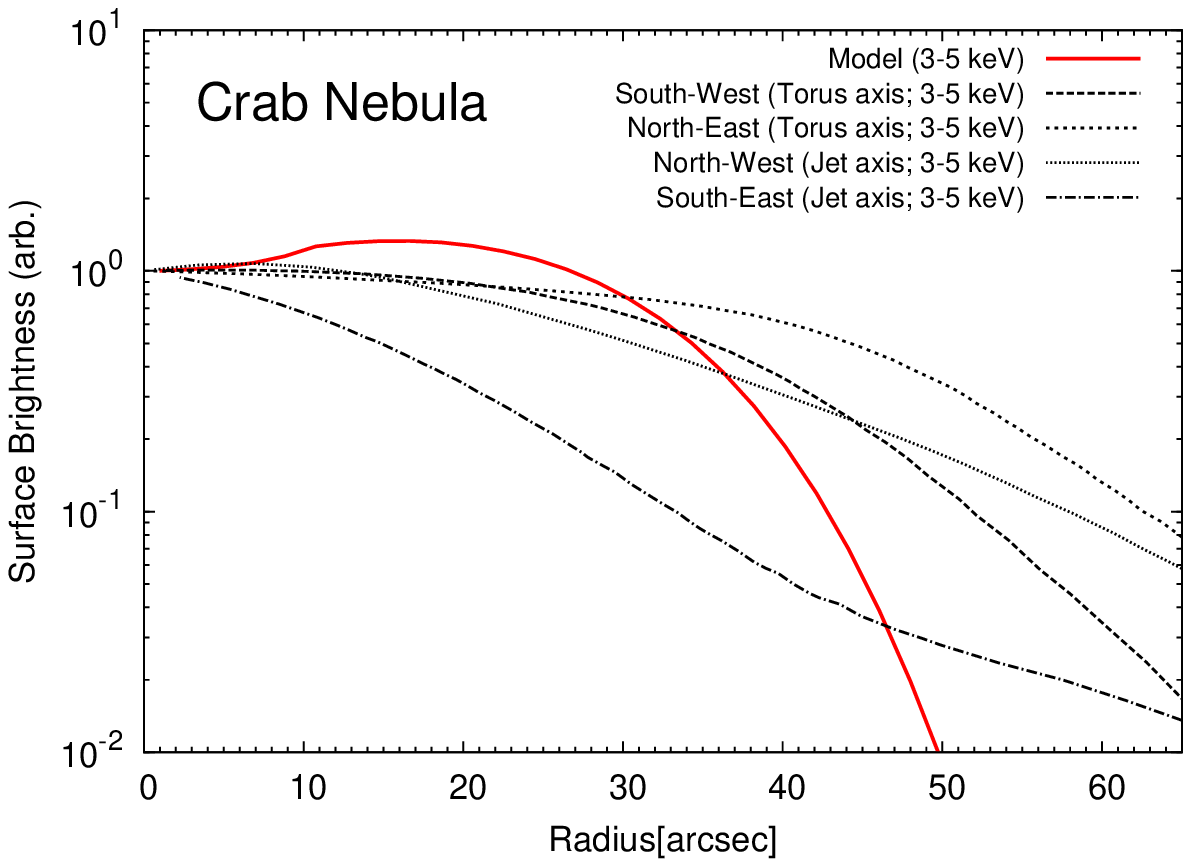}
		\end{array}$
	\end{center}
	\caption{
		An entire photon spectrum (left panel) and a radial profile of surface brightness in 3.0-5.0 keV range (right panel) for the Crab Nebula.
		The data points of the spectral energy density are taken from \citet{1977AA....61...99B} (radio),
		\citet{2010ApJ...711..417M} (radio, optical),
		\citet{1979PASP...91..436G,2006AJ....132.1610T,1968ApJ...152L..21N} (IR),
		\citet{2001AA...378..918K} (X-ray, $\gamma$-ray),
		and \citet{2004ApJ...614..897A,2006AA...457..899A,2010ApJ...708.1254A} (very high energy $\gamma$-ray).
		The data points of the X-ray surface brightness are taken from \citet{2015ApJ...801...66M}.
	}
	\label{fig:appendix:crab}
\end{figure*}

\begin{table*}[hbtp]
	\begin{center}
		\begin{tabular}{lcr}
			\hline\hline
			Given parameters $^{\rm a}$ &   Symbol   & Crab \\
			\hline
			Spin-down luminosity {\rm (erg~s$^{-1}$)} & $L_{\rm sd}$ & $5\times10^{38}$ \\
			Distance (kpc) & $D$ & 2.0 \\
			Radius of the nebula (pc) & $r_{\rm N}$ & 1.8 \\
			\hline
			Fitting parameters $^{\rm a}$\\
			\hline
			Break energy (eV) & $E_{\rm b}$ & $2.5\times10^{11}$ \\
			Low energy power-law index & $p_1$ & $1.6$  \\
			High energy power-law index & $p_2$ & $2.4$\\
			Radius of the termination shock (pc) & $r_{\rm s}$ & $0.1$\\
			Magnetization parameter & $\sigma$ & $5.0\times10^{-3}$ \\
			\hline
			Obtained parameters\\
			\hline
			Initial bulk Lorentz factor & $\gamma_{\rm u}$ & $3.2\times10^3$ \\
			Pre-shock density (cm$^{-3}$) & $n_{\rm u}$ & $1.7 \times 10^{-9}$ \\
			Pre-shock magnetic field ($\mu$G) & $B_{\rm u}$ & $30$ \\
			Maximum Energy (eV) & $E_{\rm max}$ & $2.7\times10^{15}$ \\
			Advection time (yr) & $t_{\rm adv}$ & $380$ \\
			Averaged magnetic field ($\mu$G) & $B_{\rm av}$ & $234$ \\
			Total pressure at $r=r_{\rm N}$ (erg~cm$^{-3}$) & $p_{\rm tot}$ & $2.0\times10^{-9}$ \\
			$r_{\rm eq} / r_{\rm N}$ &  & $0.45$\\
			\hline\hline
		\end{tabular}

	\vspace{1mm}
{\scriptsize 
	$^{\rm a}$ The parameters denoted with ``Given parameter'' and ``Fitting parameter'' are adopted the same value as AA96.
}

	\end{center}
	\caption{Parameters in the calculations for the Crab Nebula.}
	\label{Crab_Parameter}
\end{table*}

In Figure \ref{fig:appendix:crab}, the entire photon spectrum and the X-ray radial profile for the Crab Nebula are shown.
All the parameters to calculate the spectrum for the Crab Nebula are same as AA96 without ``Obtained parameter'', and are summarized in Table \ref{Crab_Parameter}.
The flux of inverse Compton becomes a little smaller than the value observed and calculated by AA96.
The difference in the SSC flux is due to the additional parameters $\kappa$ and $E_{\rm max}$ in AA96.
Even if we take the smaller $\sigma$ to enhance the SSC flux,
the maximum energy of synchrotron emission becomes much lower than the cut-off energy of observed spectrum (see Equation (\ref{eq:nucut})).
Within our conservative model assumption for the maximum energy of particles, we are not able to fit the spectrum around the cut-off of the spectrum.

The right panel of Figure \ref{fig:appendix:crab} shows the radial profile of surface brightness in 3.0-5.0 keV range.
The extent of the X-ray nebula calculated by our model is comparable with observations.
Comparing with the cases of 3C 58 and G21.5-0.9, 
we do not have so strong motivation to improve the 1-D steady model in the case of the Crab Nebula.


\end{document}